\documentclass[12pt]{iopart}

\usepackage{amssymb}
\usepackage{amsbsy}
\usepackage{graphicx}
\usepackage{bbm}
\usepackage{bbold}
\usepackage{hyperref}

\usepackage{fancyhdr}
\newcommand{\vareps}{\varepsilon}

\begin{document}

\title{Landau levels and Shubnikov-de Haas oscillations in
monolayer transition metal dichalcogenide semiconductors}

\author{Andor Korm\'anyos$^1$,  P\'eter Rakyta$^{2,3}$, and Guido Burkard$^1$}

\address{$^1$ Department of Physics, University of Konstanz, D-78464 Konstanz, Germany} 
\address{$^2$ Department of Theoretical Physics, Budapest University of Technology and Economics, H-1111 Budafoki \'ut. 8, Hungary}
\address{$^3$ MTA-BME Condensed Matter Research Group, Budapest University of Technology and Economics, H-1111 Budafoki \'ut. 8, Hungary}

\ead{andor.kormanyos@uni-konstanz.de}
\ead{guido.burkard@uni-konstanz.de}

\begin{abstract}
We study the Landau level  spectrum using  a multi-band $\mathbf{k}\cdot\mathbf{p}$ theory in
monolayer transition metal dichalcogenide semiconductors. We find  that in a wide magnetic field 
range the Landau levels can be characterized by a harmonic oscillator spectrum 
and  a linear-in-magnetic field term which describes the valley degeneracy breaking. 
The effect of the non-parabolicity of the band-dispersion on the Landau level 
spectrum is also discussed. 
Motivated by recent magnetotransport experiments, we use the self-consistent Born 
approximation and the Kubo formalism to  calculate the Shubnikov de-Haas oscillations  
of the longitudinal conductivity. 
We investigate how the doping level, the spin-splitting of the bands and  the 
broken valley degeneracy of the Landau 
levels affect  the magnetoconductance oscillations.
We consider monolayer MoS$_2$  and WSe$_2$ as concrete examples and 
compare the results of  numerical calculations and  an analytical formula which 
is valid in the semiclassical regime. 
Finally, we briefly analyze the recent experimental results (Reference \cite{hone-SdH}) 
using the theoretical approach we have developed. 
\end{abstract}

 
 


\section{Introduction}

Atomically thin transition metal dichalcogenides semiconductors (TMDCs)
\cite{nature-phys-review,chemsoc-review,kp-review} are   recognized 
as a material system  which, due to its finite band gap, may have a complementary functionality 
to graphene, the best known member of the family of atomically thin materials.
The experimental evidence that TMDCs become direct band gap materials in the monolayer 
limit \cite{heinz2010} and that the valley degree of  freedom \cite{dxiao} can be directly 
addressed by  optical means \cite{heinz2012,cui2012,cao,sallen} 
have spurred a  feverish research activity into the optical properties of these 
materials \cite{korn,jones,gedik,urbaszek2015a}. 
Equally influential has proved to be the fabrication of  transistors based on monolayer MoS${_2}$ 
\cite{kis-transistor1} which motivated a lot of subsequent research to understand the 
transport properties of these materials. 
Achieving good Ohmic contact to monolayer TMDCs is still challenging and this 
complicates the investigation of intrinsic properties through transport measurements. 
Nevertheless, significant  progress has been made recently in reducing the contact resistance by e.g., 
using local gating techniques \cite{jarillo-herrero2015}, phase engineering \cite{chhowalla},   
making use of  monolayer graphene as electrical contact 
\cite{zhixian,hone-SdH,duan-graphene-electr}, or 
selective etching procedure \cite{ningwang-SdH}.

Our main interest here  is to study magnetotransport properties of monolayer  TMDCs. 
Unfortunately, the relatively strong disorder in monolayer TMDC samples   have to-date hindered 
the observation of the quantum Hall effect.  
Nevertheless,  the classical Hall conductance has been  measured in a number of experiments
\cite{jarillo-herrero2015,hone-SdH,radisavljevic2013,jarillo-herrero2013,peide-ye}   
and was used to determine the charge density $n_e$ and to extract the  Hall mobility $\mu_H$. 
{In addition}, three recent works have  reported very promising progress in the efforts 
to uncover magnetic field induced quantum effects in monolayer TMDCs. 
Firstly, in Reference~\cite{goki-eda} 
the weak-localization effect was observed in 
monolayer MoS$_2$. Secondly,  it was shown that  in  boron-nitride encapsulated 
mono- and few layer MoS$_2$ \cite{hone-SdH}  and in few layer WSe$_2$ \cite{ningwang-SdH} 
it  was possible to measure  the Shubnikov-de Haas (SdH) oscillations  of the longitudinal 
resistance. 
Both of these developments are very significant  and can 
provide complementary informations: the weak localization corrections   about the
{coherence length}  and spin relaxation processes \cite{shen,OchoaFalko}, 
whereas SdH oscillations about the cross-sectional area of the Fermi surface 
and the effective mass of the carriers.

Here we first briefly review the most important steps to  calculate the LL spectrum 
in monolayer semiconductor TMDCs in perpendicular magnetic field using a 
multi-band $\mathbf{k}\cdot\mathbf{p}$ model\cite{kp-review}. 
We  show that for magnetic fields of $B \lesssim 20\,$T a simple approximation
can be applied to capture all the salient features of the LL spectrum. 
Motivated by recent experiments in MoS$_2$ \cite{hone-SdH} 
and WSe$_2$ \cite{ningwang-SdH}, 
we use the LL spectrum and the self-consistent Born approximation (SCBA) to
calculate the SdH oscillations  of the longitudinal conductance $\sigma_{xx}$. 
We discuss how  the intrinsic spin-orbit coupling and the valley degeneracy 
breaking (VDB) of the  magnetic field  affect  the magnetoconductance oscillations.   
We also point out the different  scenarios that can occur depending 
on the doping level.


\section{Landau levels in monolayer TMDCs}
\label{sec:LL-in-kp}

Electronic states in the $K$ and $-K$ valleys are related by 
time reversal symmetry in monolayer TMDCs  and hence in the presence of a 
magnetic field their degeneracy should be lifted. 
(Note that in the case of graphene the
inversion symmetry, which is present there but not in monolayer TMDCs, 
ensures that in the non-interacting limit the  LLs remain degenerate in 
the $K$ and $-K$ valleys.)
Recently several works have calculated the Landau level (LL) spectrum of 
monolayer TMDCs using the tight-binding (TB) method \cite{chuanwei-zhang,ho,asgari2015} 
and found  that the magnetic field can indeed  lift  
the degeneracy of the LLs in different valleys.
However, due to the relatively large  number of atomic orbitals 
that is needed to capture the zero magnetic field band structure, for certain problems,  
such as the SdH oscillations of longitudinal conductance, the 
TB methodology  does not offer a   convenient starting point. 
On the other hand, a simplified two-band $\mathbf{k}\cdot\mathbf{p}$ model 
was used to predict unconventional quantum Hall effect \cite{niu2013a} and to  discuss 
valley polarization \cite{niu2013b} and magneto-optical properties \cite{rose}.  
This model, however, did not capture the valley degeneracy breaking and was therefore 
in contradiction with the TB results and the considerations based on symmetry arguments. 

We first show that the VDB 
in perpendicular magnetic field can be described by  starting from a more general, 
seven-bands $\mathbf{k}\cdot\mathbf{p}$ model \cite{kp-review}.  
To this end we introduce an extended two-band continuum model which  
can be easily compared to previous works \cite{niu2013a,niu2013b,rose,asgari2013}. 
We then show  a relatively simple approximation for the LL energies which  
will prove to be useful for the calculation of  the SdH oscillations  in Section \ref{sec:SdH-osc}.

\subsection{LLs from an extended two-band model}
\label{subsec:exact-qm}

Our starting point to discuss the magnetic field effects in monolayer TMDCs is a
seven-band  $\mathbf{k}\cdot\mathbf{p}$  model  
(fourteen-band,  if the spin degree is also taken into account), we refer the reader to  
Refs.~\cite{kp-review} for details. 
In order to take into account the effects of a perpendicular magnetic field, one may use the 
Kohn-Luttinger prescription, i.e.,  we replace  the wavenumbers $\mathbf{q}=(q_x,q_y)$  appearing 
in the seven-band model with operators: 
$\mathbf{q}\rightarrow \hat{\mathbf{q}}=\frac{1}{i}\boldsymbol{\nabla}+\frac{e}{\hbar}\mathbf{A}$, 
where 
$\mathbf{A}^{T}=(0, B_z x, 0)
$ 
is the vector potential in Landau gauge and   $e>0$ is the magnitude of the electron charge. 
Note that due to this replacement 
$\hat{q}_{+}=\hat{q}_x+i \hat{q}_y $ and  $\hat{q}_- = \hat{q}_x-i \hat{q}_y $ 
become non-commuting operators: 
$
[\hat{q}_-,\hat{q}_+] =\frac{2 e B_z }{\hbar},
$
where $|B_z|$ is the strength of the magnetic field and 
$[\dots]$ denotes the commutator.
Working with a seven-band model is  not very convenient and therefore one 
may want to obtain an effective model that involves fewer bands. This can be 
done using L\"owdin-partitioning to project out those degrees of freedom 
from the seven-band Hamiltonian that are far from the Fermi energy. 
Since $\hat{q}_{+}$ and $\hat{q}_-$ are non-commuting operators, 
it is  important to keep  their order  when one performs the L\"owdin-partitioning.   
To illustrate this point  we first consider   a two-band model (four-band including spin) 
which involves the valence and the conduction bands (VB and CB). We will follow the notation 
used in Reference \cite{kp-review}. 
One finds that the low-energy effective Hamiltonian in a perpendicular magnetic field is  given by
\begin{equation}
 H_{\rm eff}^{\tau,s}=H_0+H_{\rm so}^{\tau,s}+H_{\mathbf{k}\cdot\mathbf{p}}^{\tau,s}
 \label{eq:full-eff-Ham-at-K}
\end{equation}
where $s=1$  ($s=-1$) denotes spin $\uparrow$ ($\downarrow$) and 
\begin{eqnarray}
H_0=\frac{\hbar^2}{2 m_e} \frac{\hat{q}_+\hat{q}_-+\hat{q}_-\hat{q}_+}{2}+ 
\frac{1}{2} g_e \mu_B B_z s_z
\label{eq:free-electron}
\end{eqnarray}
is the free electron term ($g_e\approx 2$ is the g-factor and $\mu_B$ is the Bohr magneton). 
Furthermore, 
\begin{eqnarray}
 H_{\rm so}^{\tau,s}=
 \left(
 \begin{array}{cc}
     \tau \Delta_{\rm vb} s_z & 0\\
  0 &  \tau \Delta_{\rm cb} s_z 
 \end{array}
 \right)
 \label{eq:H_so-K}
\end{eqnarray}
describes the spin-orbit coupling in VB and CB ($s_z$ is a spin Pauli matrix) and  
$\tau=\pm 1$ for the $\pm K$ valleys. 
The $\mathbf{k}\cdot\mathbf{p}$ Hamiltonian $H_{\mathbf{k}\cdot\mathbf{p}}^{\tau,s}$ reads
\begin{eqnarray}
H_{\mathbf{k}\cdot\mathbf{p}}^{\tau,s}=H_{\rm D}^{\tau,s}+
 H_{\rm as}^{\tau,s}+H_{3 w}^{\tau,s}+H_{\rm cub}^{\tau,s},
 \label{eq:Hkp-K}
 \end{eqnarray}
 where 
 \numparts
 \begin{eqnarray}
 H^{\tau,s}_{D}=
 \left(
 \begin{array}{cc}
  \varepsilon_{\rm vb}  & \tau\,\gamma_{\tau,s} \hat{q}_{-}^{\tau}\\
  \tau\,\gamma_{\tau,s}^{*} \hat{q}_{+}^{\tau}&  \varepsilon_{\rm cb} 
 \end{array}
 \right), \label{eq:2dDirac}\\
H^{\tau, s}_{\rm as}=
 \left(
 \begin{array}{cc}
   \alpha_{\tau,s} \hat{q}_{+}^{\tau} \hat{q}_{-}^{\tau} & 0\\
   0 &   \beta_{\tau,s}^{} \hat{q}_{-}^{\tau}\hat{q}_{+}^{\tau} 
 \end{array}
 \right), \label{eq:kp-asym}\\
 H^{\tau, s}_{3w}= 
 \left(
 \begin{array}{cc}
   0 & \kappa_{\tau,s} (\hat{q}_{+}^{\tau})^{2} \\
   \kappa_{\tau,s}^{*} (\hat{q}_{-}^{\tau})^{2} & 0
 \end{array}
 \right),\label{eq:kp-3w}\\
 H^{\tau,s}_{cub,1}=-\frac{\tau}{2}
 \left(
 \begin{array}{cc}
   0 &  \eta_{\tau,s}^{(1)} \hat{q}_{+}^{\tau} \hat{q}_{-}^{\tau}  \hat{q}_{-}^{\tau}+ 
   \eta_{\tau,s}^{(2)}  \hat{q}_{-}^{\tau} \hat{q}_{-}^{\tau}  \hat{q}_{+}^{\tau} \\
     (\eta_{\tau,s}^{(1)})^{*} \hat{q}_{+}^{\tau} \hat{q}_{+}^{\tau} \hat{q}_{-}^{\tau}+ 
   (\eta_{\tau,s}^{(2)})^{*}  \hat{q}_{-}^{\tau} \hat{q}_{+}^{\tau}  \hat{q}_{+}^{\tau} & 0
 \end{array}
 \right)\label{eq:kp-cub1}.
\end{eqnarray}
\endnumparts
Here the operator $\hat{q}_{\pm}^{\tau}$ is defined as 
$\hat{q}_{\pm}^{\tau}=\hat{q}_x\pm i \tau \hat{q}_y$. 
The material specific properties are encoded in the parameters 
${\vareps}_{\rm vb}$, ${\vareps}_{\rm cb}$ 
(band-edge energies in the absence of SOC),  $\gamma_{\tau,s}$ 
(coupling between the VB and the CB) and 
$\alpha_{\tau,s}$, $\beta_{\tau,s}$, $\kappa_{\tau,s}$, $\eta_{\tau,s}^{(1)}$,  
$\eta_{\tau,s}^{(2)}$, 
which describe the effects of virtual transitions between the VB (CB) and the other bands in the 
seven-band model. 
In general, the off-diagonal material parameters $\gamma_{s,\tau}$, $\kappa_{s,\tau}$ and 
$\eta_{s,\tau}^{(1)}$, $\eta_{s,\tau}^{(2)}$ are complex numbers such that for the $-K$ 
valley ($\tau=-1$) they are the complex conjugates of the $K$ valley case ($\tau=1$).
In the absence of a magnetic field, the material parameters appearing in  
Eqs.~\eref{eq:2dDirac} - \eref{eq:kp-cub1} can be obtained by, e.g., fitting the 
eigenvalues of $ H_{\rm eff}^{\tau,s}$  to  the band structure obtained from density functional 
theory (DFT) calculations. We refer to Reference~\cite{kp-review} for the details of 
this fitting procedure and for tables containing the extracted parameters for 
monolayer semiconductor TMDCs. 
Here we only  mention that such a fitting procedure yields real numbers which  
depend on the spin index $s$ but do not depend explicitly on the valley index $\tau$. 
(The parameters $\eta_{\tau,s}^{(1)}$ and $\eta_{\tau,s}^{(2)}$ cannot be obtained
 separately from fitting the DFT band structure, only their sum, 
 $\eta_{\tau,s}^{}$ can be extracted. Fortunately, as we will see below,
 the effect of $H^{\tau,s}_{cub,1}$ is very small in the magnetic 
 field range we are primarily interested in. )

We note that a $\mathbf{k}\cdot\mathbf{p}$ model, similar to ours, was recently 
used in References~\cite{asgari2015,asgari2013} to calculate the LL spectrum.   
{There are two  differences between our $\mathbf{k}\cdot\mathbf{p}$
Hamiltonian Eqs.~\eref{eq:Hkp-K} and the model in  
References~\cite{asgari2015,asgari2013}. The first one is that  
higher order terms that would correspond to our 
$H^{\tau, s}_{3w}$ and $H^{\tau,s}_{cub,1}$ were not considered in 
References~\cite{asgari2015,asgari2013}. We  keep these terms in order 
to see more clearly the  magnetic field range where the approximation discussed
in Section \ref{subsec:LL-approx} is valid. The second difference can be found
in our $H^{\tau, s}_{\rm as}$ \eref{eq:kp-asym} and 
the corresponding  Hamiltonian  used in \cite{asgari2015,asgari2013}.
}
This difference can be traced back to the way  the magnetic field is taken into account 
in the effective models that are obtained from multi-band Hamiltonians. 
In References \cite{asgari2015,asgari2013} 
first an effective zero field two-band model was derived and then in a second step the 
Luttinger-prescription was performed in this effective model.
Therefore the terms which are $\sim q^2$ 
in the zero field case become $\sim \hat{q}_{+}\hat{q}_{-}+\hat{q}_{-}\hat{q}_{+}$ after 
the  Luttinger-prescription.
In contrast, as mentioned above, we  perform the Luttinger prescription in the multi-band 
Hamiltonian and obtain the effective two-band model 
$ H_{\rm eff}^{\tau,s}$ \eref{eq:full-eff-Ham-at-K} 
in the second step.  The two approaches may lead to different results because the operators 
$\hat{q}_+$, $\hat{q}_-$ do not commute and this  should be taken into account 
in the L\"owdin-partitioning which yields  the effective two-band model.

\begin{figure}[ht]
\begin{center}
 \includegraphics[scale=0.55]{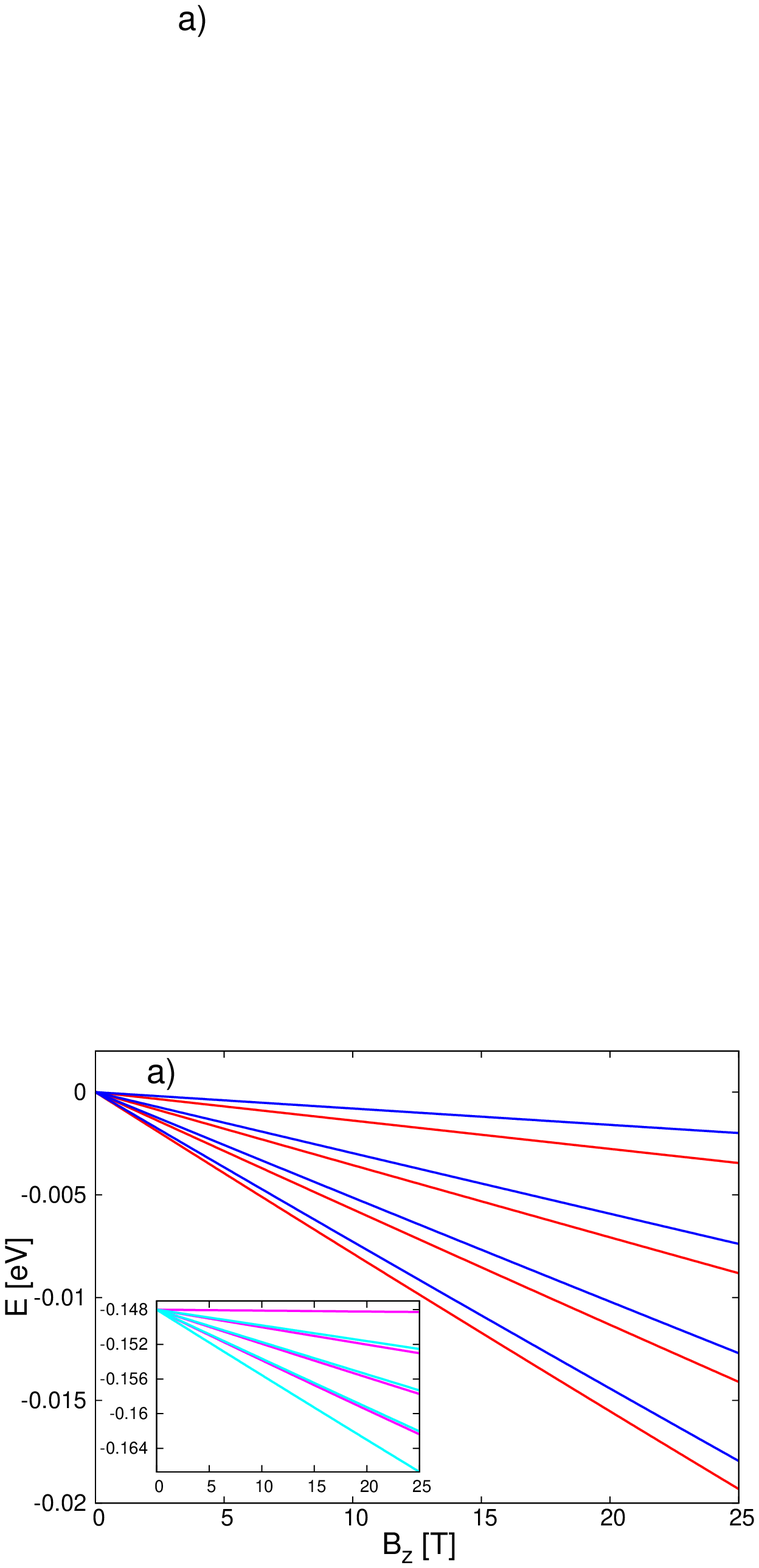}\,\,\,\,\,\,
 \includegraphics[scale=0.55]{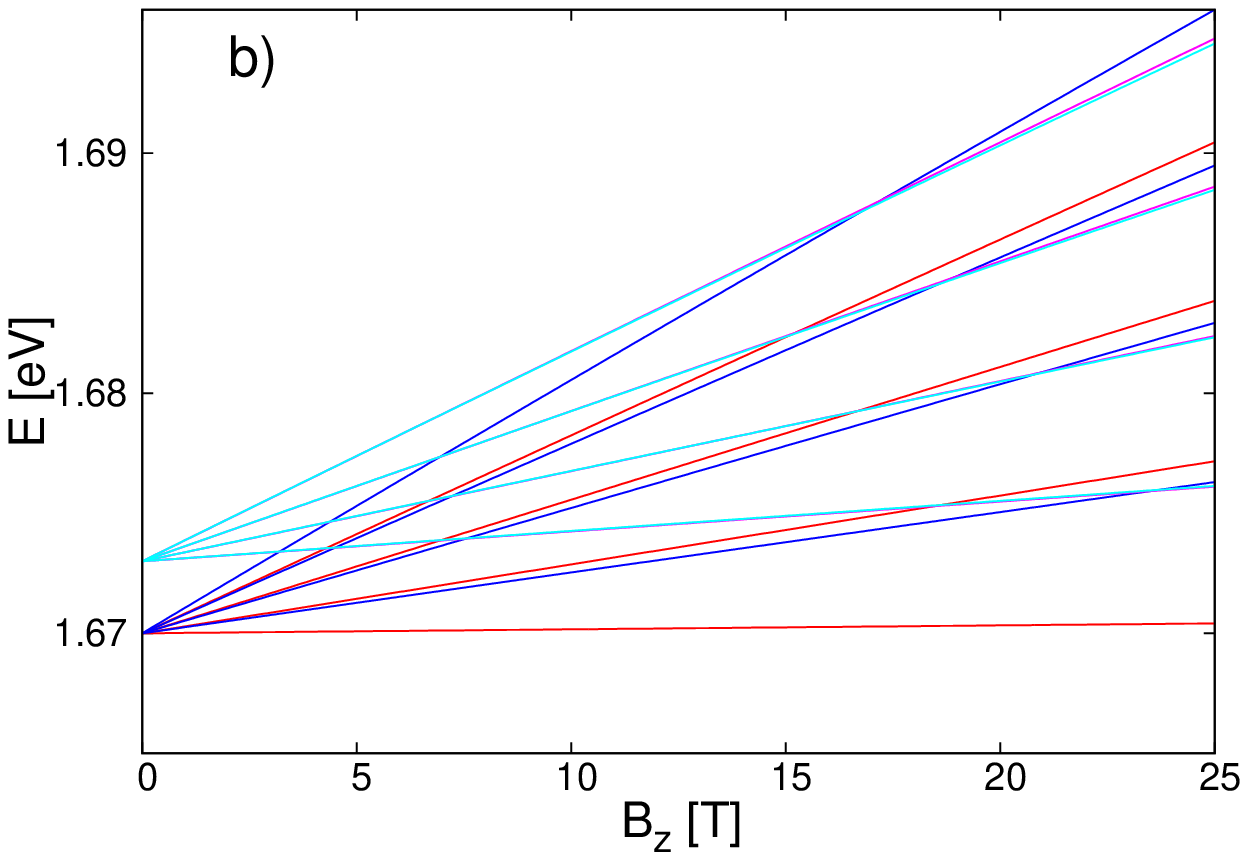}
 \end{center}
 \caption{ Numerically calculated LL spectrum of MoS$_2$. 
  a) The first few  LL in the higher spin-split VB. Red lines: the $K$ valley ($\tau=1$), 
   blue lines:  the $-K$ valley ($\tau=-1$). The inset shows the LLs in the lower spin-split VB.  
  b)  The first few  LL in the CB. LLs both in lower spin-split band and in the higher spin-split 
      band are shown. Red and purple lines: the $K$ valley, blue and cyan:the $-K$ valley. } 
 \label{fig:LLs-MoS2-exact}
\end{figure}

\begin{figure}[ht]
\begin{center}
\includegraphics[scale=0.55]{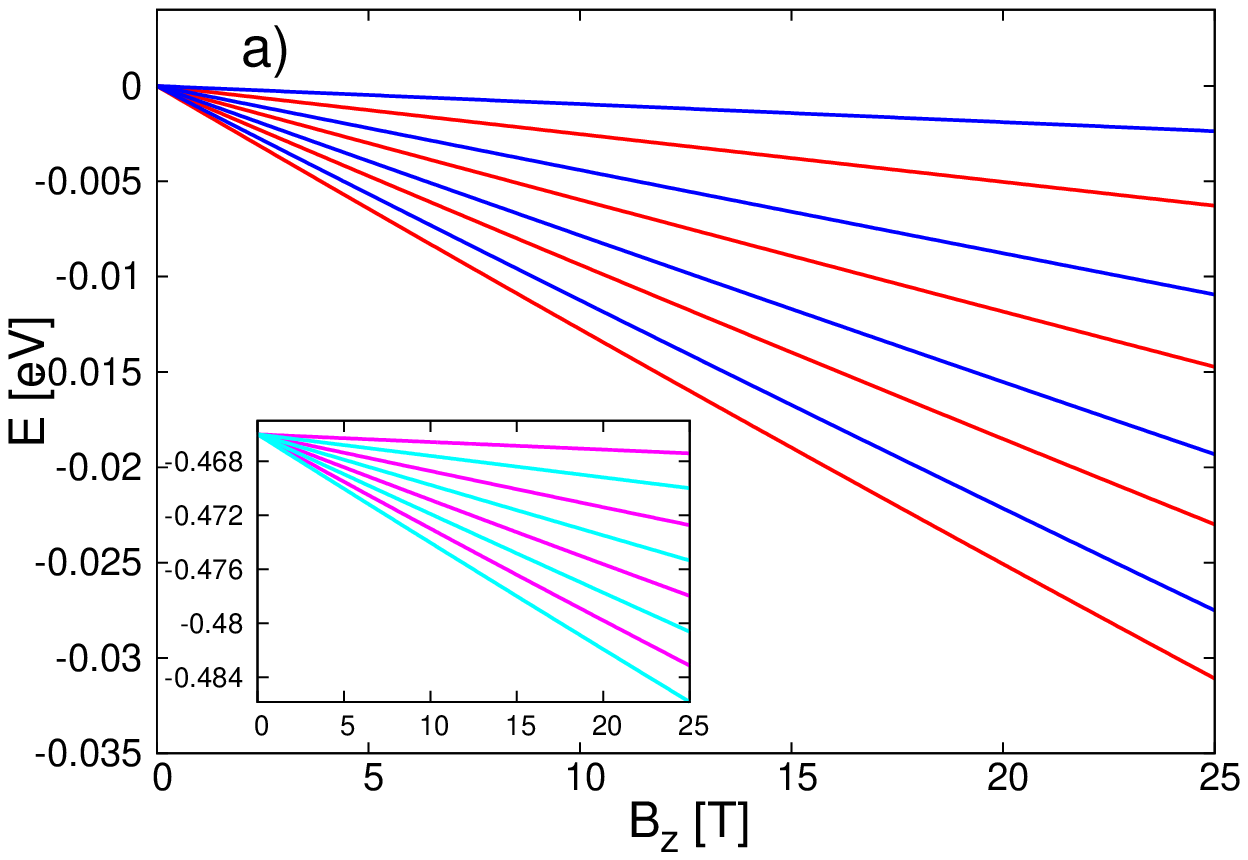}\,\,\,\,\,\,
 \includegraphics[scale=0.55]{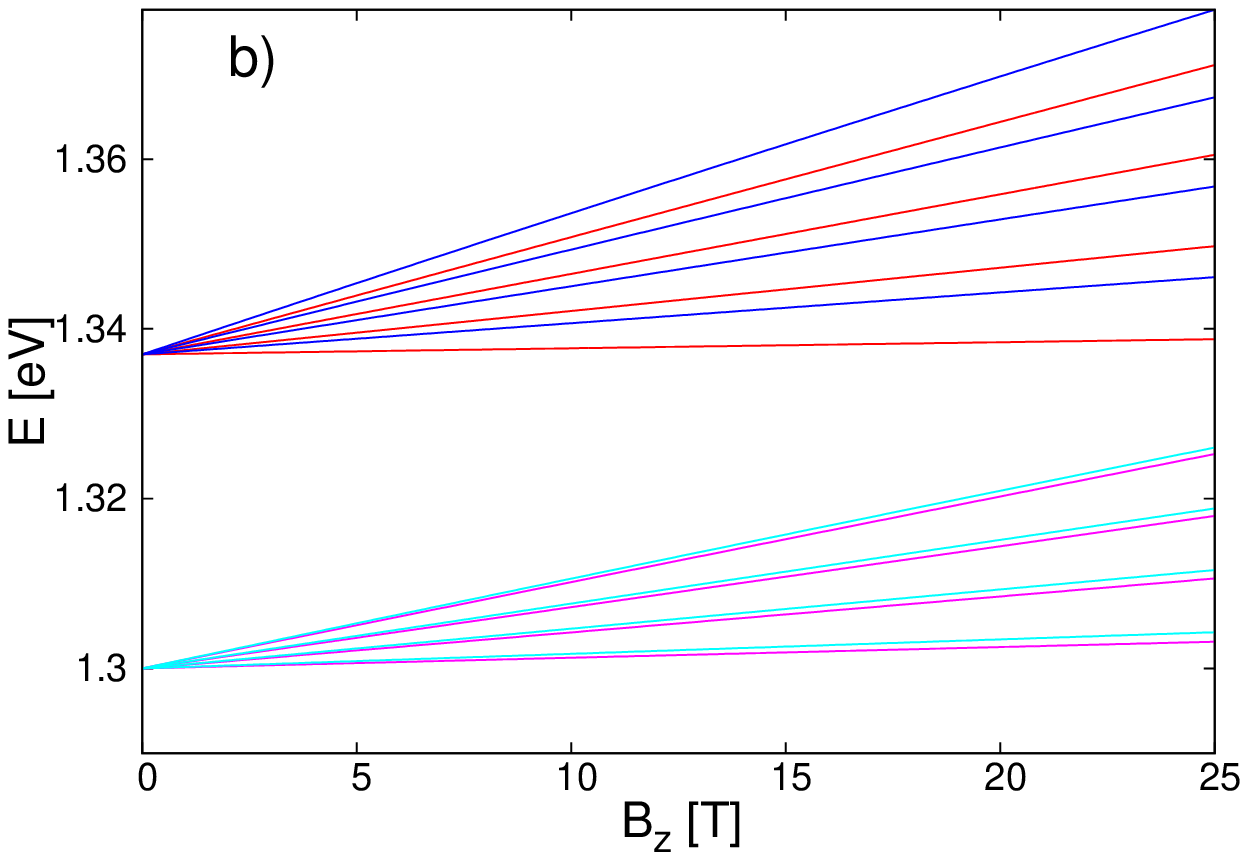}
 \end{center}
 \caption{ Numerically calculated LL spectrum of WSe$_2$. 
  a) The first few  LL in the higher spin-split VB. Red lines: the $K$ valley ($\tau=1$), 
   blue lines:  the $-K$ valley ($\tau=-1$). The inset shows the LLs in the lower spin-split VB.  
  b)  The first few  LL in the CB. LLs both in lower spin-split band and in the higher spin-split 
      band are shown. Red and purple lines: the $K$ valley, blue and cyan:the $-K$ valley. } 
 \label{fig:LLs-WSe2-exact}
\end{figure}

The spectrum of  $ H_{\rm eff}^{\tau,s}$  can be calculated numerically  
using harmonic oscillator eigenfunctions as basis states.  Taking $B_z>0$  
for concreteness, one can see that 
the operators $a$ and $a^{\dagger}$ defined as 
$\hat{q}_-=\frac{\sqrt{2}}{l_b} a$, $\hat{q}_+=\frac{\sqrt{2}}{l_b} a^{\dagger}$,
where $l_B=\sqrt{\hbar/(e |B_z|)}$,  satisfy the bosonic commutation 
relation $[a,a^{\dagger}]=1$. 
(For $B_z<0$ one has to define  
$\hat{q}_+=\frac{\sqrt{2}}{l_b} a$, $\hat{q}_-=\frac{\sqrt{2}}{l_b} a^{\dagger}$). 
Therefore  one  can  calculate the matrix elements of  $ H_{\rm eff}^{\tau,s}$ 
in a large, but finite harmonic oscillator basis
and  diagonalize the resulting matrix. For a large enough number of basis states  the lowest 
eigenvalues  of  $ H_{\rm eff}^{\tau,s}$  will not depend on the 
exact number of the basis states. 
Such a LL calculation is shown in  Figure~\ref{fig:LLs-MoS2-exact} for MoS$_2$  
and in   Figure~\ref{fig:LLs-WSe2-exact}  for WSe$_2$  (we have used 
the material parameters given in Reference \cite{kp-review}).
One can see that the LLs in different valleys are not degenerate and that the 
magnitude of the valley degeneracy breaking is different in the VB and CB 
and for the lower and higher-in-energy  spin-split bands. While the results in the VB are 
qualitatively similar for MoS$_2$ and WSe$_2$, considering the CB,   
for MoS$_2$ the valley splitting of the LLs is smaller in the higher spin-split
band, whereas the opposite is true for WSe$_2$. This is a consequence of the interplay of the 
Zeeman term in Eq.~\eref{eq:free-electron} and other, band-structure related terms which 
lead to VDB. 
(For MoS$_2$ the valley splitting in the higher spin-split CB (purple and cyan lines) is very small 
for the material parameter set used in these calculations and can only be 
noticed for large magnetic fields.) 
One can also observe  that in the CB the lowest LL is in  valley  $K$, 
whereas in the VB  it is in valley $-K$.

Further details of the VDB, including its  dependence on the  
parameter set that can be extracted from DFT calculations, 
will be discussed  in Section \ref{subsec:LL-approx}. 
Here we point out that these results qualitatively 
agree with the TB calculations of Reference \cite{chuanwei-zhang,ho,asgari2015}, i.e., the 
continuum approach can reproduce all important features of multi-band TB calculations. 
A more quantitative comparison between our results and the TB results  
\cite{chuanwei-zhang,ho,asgari2015}
is difficult, partly because the details may depend on the way how the material parameters 
are extracted from the DFT band structure and also because in the TB calculations the 
Zeeman effect was often neglected.   

The LL energies  can also be obtained analytically in the approximation where
$H^{\tau, s}_{3w}$ and  $H^{\tau,s}_{cub,1}$ are neglected.  
We will not show these analytical results here because it turns out that an 
even simpler approximation 
yields a good agreement with the numerical calculations 
shown in Figures \ref{fig:LLs-MoS2-exact} and 
\ref{fig:LLs-WSe2-exact} (see Section \ref{subsec:LL-approx}) and offers a 
suitable starting point 
to develop a theory for the SdH oscillations of the longitudinal conductivity.


\subsection{Approximation of the LLs spectrum }
\label{subsec:LL-approx}

In zero magnetic field,  the trigonal warping term Eq.~\eref{eq:kp-3w} 
and the third order term Eq.~\eref{eq:kp-cub1} are important in order to understand the results
of recent angle resolved photoelectron spectroscopy measurements and in order to obtain 
a good fit to the DFT band structure, respectively \cite{kp-review}.
However, as we will show   for the calculation of LLs  the terms  $H^{\tau, s}_{3w}$  and 
$H^{\tau,s}_{cub,1}$ are less important. 
To see this one can perform  another L\"owdin-partitioning on $H_{\rm eff}^{\tau,s}$ 
to obtain  effective singe-band Hamiltonians for 
the VB and the CB separately. Keeping only lowest order terms in $B_z$ one finds that 
these single-band Hamiltonians correspond to a harmonic oscillator Hamiltonian 
(with different effective masses in the VB and CB and for the  spin-split bands) 
and a term which describes a linear-in-$B_z$ splitting of the  energies of the 
LLs in the two valleys. 
Therefore the LL spectrum can be approximated by 
\numparts
\begin{eqnarray}
 E_{n,{\rm vb}}^{\tau,s}=\vareps_{\rm vb}^{\tau,s}+\hbar \omega_{{\rm vb}}^{(\tau,s)}
 \left(n+\frac{1}{2}\right)+\frac{1}{2} g_e \mu_B B_z s
 +\frac{1}{2} g_{vl, {\rm vb} }^{(s)} \mu_B B_z\, \tau, 
 \label{eq:single-band-LL-VB}\\
  E_{n,{\rm cb}}^{\tau,s}=\vareps_{\rm cb}^{\tau,s}+\hbar \omega_{{\rm cb}}^{(\tau,s)}
  \left(n+\frac{1}{2}\right)+\frac{1}{2} g_e \mu_B B_z s 
 + \frac{1}{2} g_{vl,{\rm cb}}^{(s)} \mu_B B_z\, \tau.  
 \label{eq:single-band-LL-CB}
\end{eqnarray}
\endnumparts
Here, the following notations are introduced: $n=0,1,2,\dots$ is an integer denoting the 
LL index, 
$\vareps_{\rm vb (cb)}^{\tau,s}=\vareps_{\rm vb (cb)}+\tau \Delta_{\rm vb (cb)} s_z$ 
are the band edge 
energies in the VB (CB) for a given spin-split band $s$ and    
$\omega_{{\rm vb\, (cb)}}^{(\tau,s)}=\frac{e B_z}{m_{\rm vb\, (cb)}^{(\tau,s)}}$  
are cyclotron frequencies. 
In terms of the parameters appearing in Eqs.~\eref{eq:free-electron}-\eref{eq:Hkp-K}, 
for $\tau=1$ the effective 
masses $m_{\rm vb\, (cb)}^{(s)}$ that enter the expression of the cyclotron frequencies 
are given by \cite{kp-review}
\numparts
\begin{eqnarray}
\frac{\hbar^2}{2 m_{\rm vb}^{(1,s)}} =\left(\frac{\hbar^2}{2 m_e}+\alpha_{s}-
\frac{|\gamma|^2}{E_{\rm bg}^{(s)}}\right)
\label{eq:effmass-VB}\\
\frac{\hbar^2}{2 m_{\rm cb}^{(1,s)}} =\left(\frac{\hbar^2}{2 m_e}+\beta_{s}+
\frac{|\gamma|^2}{E_{\rm bg}^{(s)}}\right)
\label{eq:effmass-CB}
\end{eqnarray}
\endnumparts
where ${E_{\rm bg}^{(s)}}=\vareps_{\rm cb}^{1,s}-\vareps_{\rm vb}^{1,s}$.  
The corresponding expressions for 
$\tau=-1$  can be easily found  from the requirement electronic states that are 
connected by time reversal
symmetry have the same effective mass. This means that bands corresponding to the 
same value of the product $\tau\, s$ have the same effective mass. 
The third term in Eqs~\eref{eq:single-band-LL-VB} and \eref{eq:single-band-LL-CB}
comes from the free-electron term \eref{eq:free-electron}. 
The VDB 
is described by the last term in 
Eqs.~\eref{eq:single-band-LL-VB}, \eref{eq:single-band-LL-CB} and the valley g-factors are given by 
\numparts
\begin{eqnarray}
g_{vl, {\rm vb}}^{(s)}=4\frac{m_e}{\hbar^2}\left(\alpha_{s}+\frac{|\gamma|^2}{E_{bg}^{(s)}}\right)
\label{eq:g-VB}\\ 
g_{vl, {\rm cb}}^{(s)}=4\frac{m_e}{\hbar^2}\left(\frac{|\gamma|^2}{E_{bg}^{(s)}}-\beta_{s}\right).
\label{eq:g-CB}
\end{eqnarray}
\endnumparts
As one can see from \eref{eq:g-VB}-\eref{eq:g-CB}, 
$g_{vl}^{(s)}$     depends on  the (virtual) inter-band transition 
matrix elements $\alpha_s$, $\beta_s$ and $\gamma$. Due to the intrinsic spin-orbit coupling, 
the magnitude of these matrix elements is spin-dependent~\cite{kp-review}. 
{ Note, that $g_{vl}$ 
is different in the VB and the CB. This is in agreement with  numerical calculations  based on 
multi-band tight-binding models \cite{chuanwei-zhang,asgari2015}}.
For the CB, the details of the derivation  that leads to \eref{eq:single-band-LL-CB} can be 
found in \cite{kormanyos-prx}, for the VB the derivation of \eref{eq:single-band-LL-VB}  
is analogous and therefore it will not be detailed here. 
We note that in variance to Reference \cite{kormanyos-prx}, 
we do not define separately an out-of-plane spin g-factor and a spin independent valley g-factor,  
these two g-factors are merged in $g_{vl}^{(s)}$. 
The response to magnetic field also depends on the free electron Zeeman term. 
The spin-index $s$ to be used in 
the evaluation of the Zeeman term in Eqs.~\eref{eq:single-band-LL-VB}-\eref{eq:single-band-LL-CB} 
follows the spin-polarization of the given spin-split band.  
For MoS$_2$, the spin polarizations $s$ of each band  
are shown in Figure \ref{fig:doping}, other MoX$_2$ (X=\{S, Se, Te\}) monolayer TMDCs 
have the same polarization.  
For monolayer WX$_2$ TMDCs the spin polarization in the VB is the same as for the MoX$_2$, 
but in the CB 
the polarization of the lower (higher) spin-split band is the opposite \cite{kp-review}.
We are mainly interested in how the magnetic field breaks the degeneracy of those electronic states 
which are connected by time reversal in the absence of the magnetic field. 
Using Eqs.~\eref{eq:single-band-LL-VB}-\eref{eq:single-band-LL-CB}, 
the valley splitting $\delta E_{\rm cb (vb)}^{(i)}=g^{(i)}_{eff, {\rm cb\, (vb)}} \mu_B B_z$ 
of these states  can be characterized by an effective g-factor 
$g^{(i)}_{eff, {\rm cb\, (vb)}}= (g_e s \tau + g_{vl,{\rm cb\,(vb)}}^{(s)})$, 
where $i=1 (2)$ denotes the higher-in-energy (lower-in-energy) spin-split band. 
In the VB the upper index ${(1)}$ [${(2)}$] 
is equivalent to $\downarrow$ ($\uparrow$), but in the CB the  relation  depends on the specific 
material being considered because the polarisation  is different for MoX$_2$ and WX$_2$ materials. 
Taking  first the MoX$_2$ monolayers one finds that (see also Figure~\ref{fig:doping}) 
\numparts
\begin{eqnarray}
 g^{(1)}_{eff, {\rm vb}}= (-g_e  + g_{vl,{\rm vb}}^{\downarrow})\hspace{1cm} 
 g^{(2)}_{eff, {\rm vb}}= (g_e + g_{vl,{\rm vb}}^{\uparrow})
 \label{eq:geff-VB}\\
 g^{(1)}_{eff, {\rm cb}}= (g_e  + g_{vl,{\rm cb}}^{\uparrow}) \hspace{1cm} 
 g^{(2)}_{eff, {\rm cb}}= (-g_e + g_{vl,{\rm cb}}^{\downarrow})
\end{eqnarray}
\endnumparts
For WX$_2$ monolayers  $g^{(i)}_{eff, {\rm vb}}$ can also be calculated by \eref{eq:geff-VB}, 
whereas in the CB 
\begin{eqnarray}
 g^{(1)}_{eff, {\rm cb}}= (-g_e  + g_{vl,{\rm cb}}^{\downarrow}) \hspace{1cm} 
 g^{(2)}_{eff, {\rm cb}}= (g_e + g_{vl,{\rm cb}}^{\uparrow})
\end{eqnarray}
As an example the numerical values of the various g-factors defined above are given in 
Table \ref{tbl:g-factors-MoS2} for MoS$_2$ and in Table \ref{tbl:g-factors-WSe2} for WSe$_2$.
One can see that $g_{vl,{\rm cb\,(vb)}}^{(s)}$ can be comparable in magnitude to  $g_e$. 
This explains why the valley splitting is very
small for MoS$_2$ in the case of the upper spin-split band in the CB 
(see Figure \ref{fig:LLs-MoS2-exact}), 
whereas the opposite is true for WSe$_2$ (Figure \ref{fig:LLs-WSe2-exact}).

As one can see from Eqs. \eref{eq:g-VB} and  \eref{eq:g-CB},   $g_{vl}^{(s)}$  
depends explicitly on the band-gap $E_{bg}^{(s)}$ of a given spin $s$. In addition, 
the parameters $\gamma$, $\alpha_{s}$ and 
$\beta_{s}$ implicitly also depend on  $E_{bg}^{(s)}$ due to the fitting procedure 
that is used to obtain 
them from DFT band structure calculations\cite{kp-review}.  
It is known that  $E_{bg}^{(s)}$ is underestimated in DFT calculations and its 
exact value at the moment is not known for most monolayer TMDCs. 
Therefore in Reference \cite{kp-review} we have obtained two sets of the $\mathbf{k}\cdot\mathbf{p}$ 
band structure parameters, the first one using   $E_{bg}^{(s)}$ from DFT  and the second one using 
$E_{bg}^{(s)}$ extracted from GW calculations.  
The calculations shown in  Figures \ref{fig:LLs-MoS2-exact} 
and \ref{fig:LLs-WSe2-exact} were obtained with the former parameter set. 
As shown in Table \ref{tbl:g-factors-MoS2}, the calculated g-factors depend quite 
significantly on the choice of the parameter set. 
While there is an uncertainty regarding the magnitude of $g_{vl}^{(s)}$, we expect that 
the g-factors obtained by using the DFT and the GW parameter sets will  bracket the 
actual experimental values. 
On the other hand, the effective masses are probably captured quite well by  
DFT calculations and therefore 
the first term in Eqs.~\eref{eq:single-band-LL-VB}-\eref{eq:single-band-LL-CB} 
is less affected by the uncertainties of the band structure parameters. The calculations in 
Figures \ref{fig:LLs-MoS2-exact} and \ref{fig:LLs-WSe2-exact}
correspond to the ``DFT'' parameter set 
in Tables \ref{tbl:g-factors-MoS2} and \ref{tbl:g-factors-WSe2}.  

\begin{table}[htb]
\caption{Valley g-factors in MoS$_2$. In the first row the g-factors are 
         obtained with the help of DFT band gap, in the  second row the g-factors are calculated 
         with a band gap taken from the  $GW$ calculations.
}
\label{tbl:g-factors-MoS2}
 \begin{indented}
\lineup
\item[]
\begin{tabular}{@{}lrrrrrrrrrr}
\br
    & $E_{\rm bg}^{\downarrow}$ & $E_{\rm bg}^{\rm \uparrow}$ & $g_{vl,{\rm vb}}^{\downarrow}$ & $g_{vl,{\rm vb}}^{\uparrow}$  & 
    $g_{eff, \rm{vb}}^{(1)}$ &  $g_{eff, \rm{vb}}^{(2)}$ & $g_{vl,{\rm cb}}^{\downarrow}$  & $g_{vl,{\rm cb}}^{\uparrow}$ & 
    $g_{eff, \rm{cb}}^{(1)}$ & $g_{eff, \rm{cb}}^{(2)}$\\
\br

 DFT  & $1.66\,{\rm eV}^{a}$ & $1.838\,{\rm eV}^{a}$ & $0.98$ & $0.96$ & $-1.02$ & $2.96$ & $-2.11$ & $-2.05$ &  $-0.05$ & $-4.11$\\
\mr 
 GW   &  $2.8\,{\rm eV}^{b}$  &  $2.978\,{\rm eV}^{b}$ & $2.57$ & $2.38$ & $0.57$ &  $4.38$ & $-0.52$ &  $-0.6$ &   $1.4$  &$-2.52$\\  
\br
\end{tabular}
\item[] $^{a}$ {adapted from Reference \cite{kp-review}.}
\item[] $^{b}$ {adapted from Reference \cite{qiu}.}
\end{indented}
\end{table}

\begin{table}[htb]
\caption{Valley g-factors in WSe$_2$. In the first row the g-factors are 
         obtained with the help of DFT band gap, in the  second row the g-factors are calculated 
         with a band gap taken from the  $GW$ calculations. 
}
\label{tbl:g-factors-WSe2}
 \begin{indented}
\lineup
\item[]
\begin{tabular}{@{}lrrrrrrrrrr}
\br
    & $E_{\rm bg}^{\downarrow}$ & $E_{\rm bg}^{\rm \uparrow}$ & $g_{vl,{\rm vb}}^{\downarrow}$ & $g_{vl,{\rm vb}}^{\uparrow}$  & 
    $g_{eff, \rm{vb}}^{(1)}$ &  $g_{eff, \rm{vb}}^{(2)}$ & $g_{vl,{\rm cb}}^{\downarrow}$  & $g_{vl,{\rm cb}}^{\uparrow}$ & 
    $g_{eff, \rm{cb}}^{(1)}$ & $g_{eff, \rm{cb}}^{(2)}$\\
\br

 DFT  & $1.337{\rm eV}^{a}$ & $1.766{\rm eV}^{a}$ & $-0.38$ & $-0.23$ & $-2.38$ & $1.77$ & $-2.71$ & $-2.81$ &  $-4.71$ & $-0.81$\\
\mr 
 GW   &  $2.457{\rm eV}^{b}$&  $2.886{\rm eV}^{b}$ & $2.55$ & $1.9$ &  $0.55$ &  $3.9$   & $-0.67$ &  $0.13$ & $-2.67$  & $2.13$\\  
\br
\end{tabular}
\item[] $^{a}$ {adapted from Reference \cite{kp-review}.}
\item[] $^{b}$ {adapted from Reference \cite{ashwin}.}
\end{indented}
\end{table}

\begin{figure}[ht]
\begin{center}
 \includegraphics[scale=0.55]{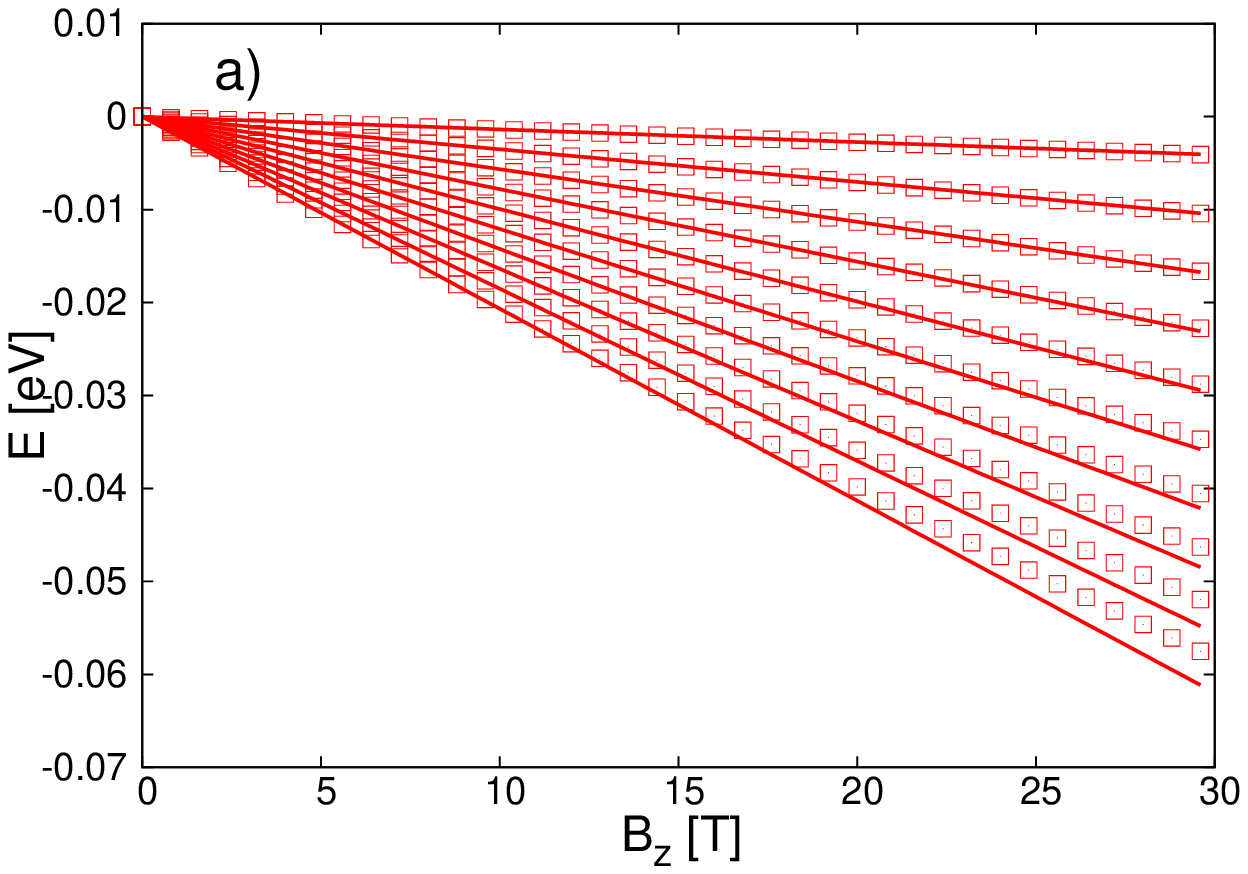}\,\,\,\,\,\,
 \includegraphics[scale=0.55]{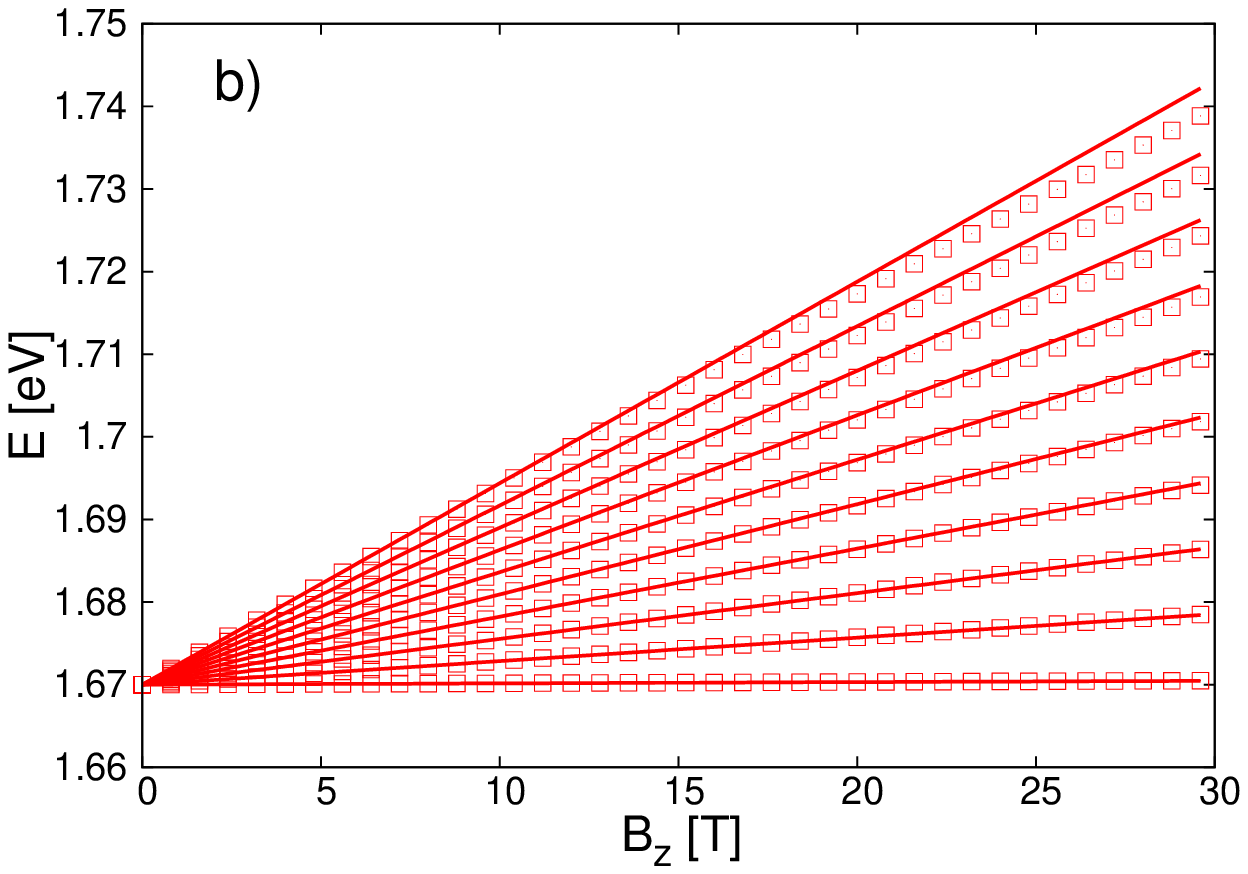}
 \end{center}
 \caption{Comparison of the LL spectrum in  MoS$_2$ obtained from the two-band 
 model and from the single band model. 
  a) The numerically calculated LLs using \eref{eq:full-eff-Ham-at-K} 
   for the $\tau=1$, $s=-1$ in the VB (squares) and the approximation 
  \eref{eq:single-band-LL-VB} (solid lines) for LL indices $n=0\dots 9$. 
   b) the same as in a) but for the for the $\tau=1$, $s=-1$ band in the CB
  (squares) and the approximation \eref{eq:single-band-LL-CB} (solid lines).} 
 \label{fig:LL-approx-MoS2}
\end{figure}

In order to see the accuracy of the approximation introduced in 
Eq.~\eref{eq:effmass-VB}-\eref{eq:effmass-CB}, 
in Figure \ref{fig:LL-approx-MoS2} we  compare the LL spectrum obtained in this approximation 
and calculated numerically using the Hamiltonian \eref{eq:full-eff-Ham-at-K}.
As one can see the approximation 
is very good both in the VB and  in the CB 
up to magnetic fields $\lesssim 20 {\rm T}$. For larger magnetic fields and large LL indices 
($n > 7$) deviations start to appear between the full quantum results and the approximation. 
The deviations are stronger in the VB which  we attribute  to the larger trigonal 
warping \cite{kp-review} of the band structure  in the VB.  To our knowledge the effects
of the non-parabolicity of the band-dispersion on the LL spectrum has not been 
discussed before for monolayer TMDCs. 

Given the noticeable  uncertainty regarding the exact values of the effective g-factors, 
one may ask which features of the LL spectrum are affected or remain qualitatively the same. 
Looking at Tables \ref{tbl:g-factors-MoS2} and \ref{tbl:g-factors-WSe2}, one can see  
that in some cases only the magnitude of an effective g-factor changes, in other cases 
both the magnitude and the sign. 
Firstly, we consider a case which illustrates possible effects of  the uncertainty in 
the magnitude of an effective g-factor. 
\begin{figure}[ht]
\begin{center}
 \includegraphics[scale=0.55]{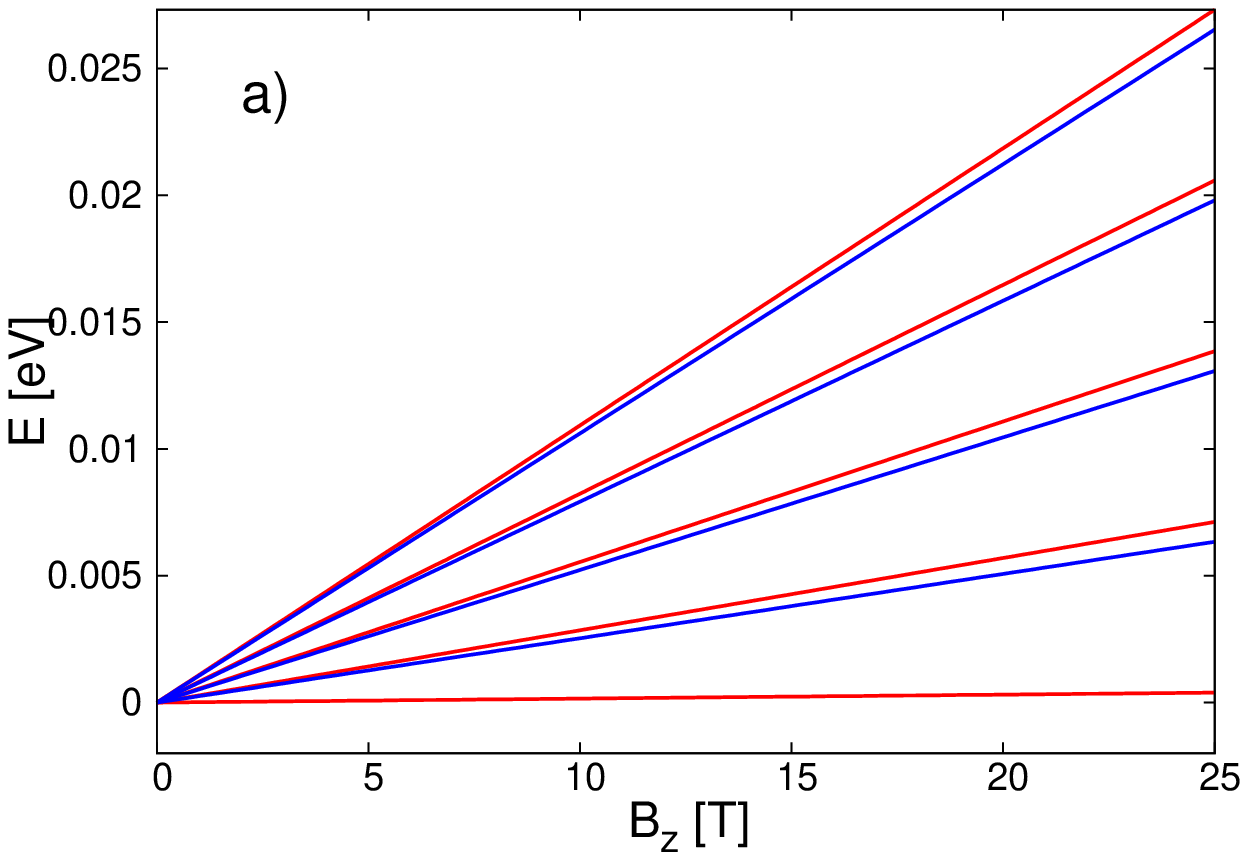}\,\,\,\,\,\,
 \includegraphics[scale=0.55]{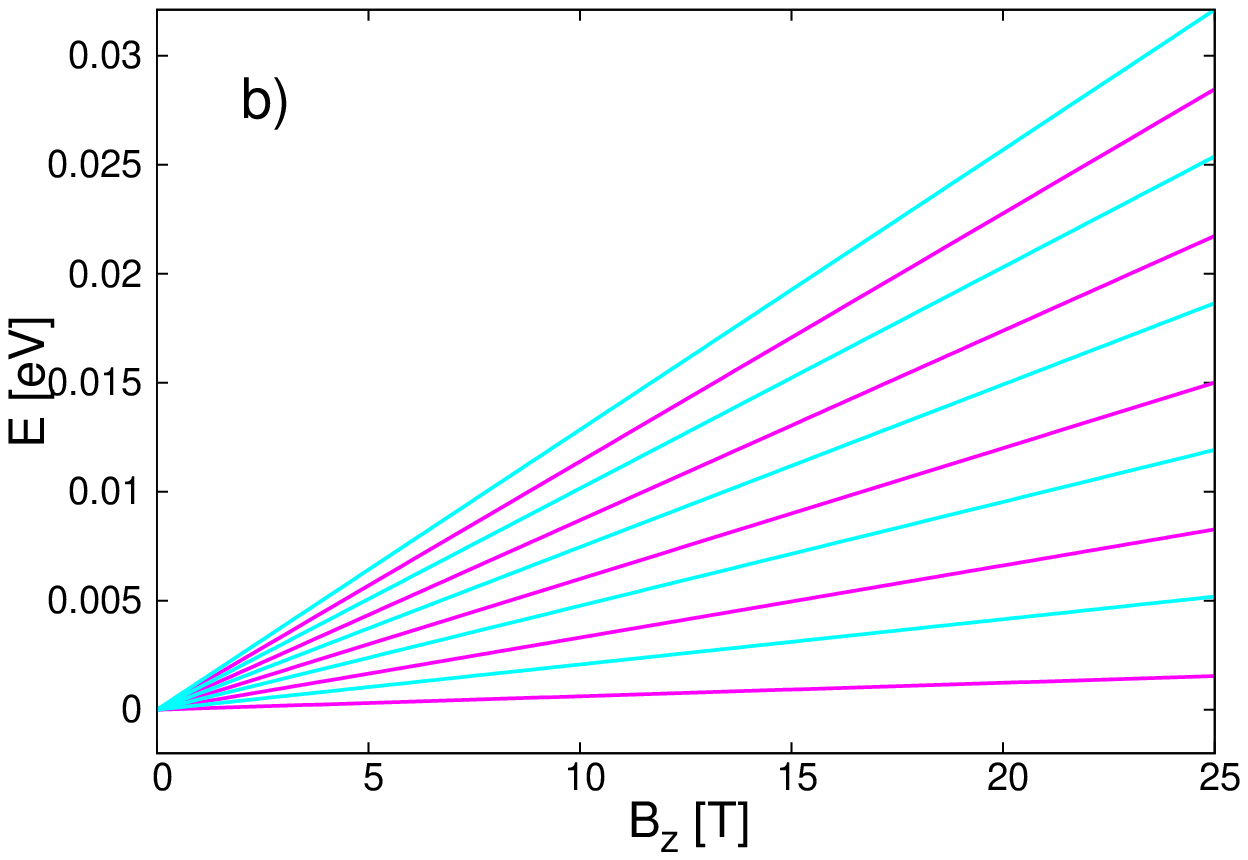}
 \end{center}
 \caption{Comparison of the LL spectrum in in the lower-in-energy spin-split CB of MoS$_2$ 
 obtained with a) $g_{eff, {\rm cb}}^{(2)}=-4.11$  and b)  $g_{eff, {\rm cb}}^{(2)}=-2.52$.
  LLs in  different valleys are denoted by different colors. } 
 \label{fig:LL-gDFTvGW}
\end{figure}
In Figure \ref{fig:LL-gDFTvGW} we show  the LLs in the lower spin-split 
CB in MoS$_2$ for the two different $g_{eff,{\rm cb}}^{(2)}$ given in Table \ref{tbl:g-factors-MoS2}. 
One can see that in  Figure \ref{fig:LL-gDFTvGW}(a) the VDB is small, 
except for the lowest LL, which is clearly separated from the other LLs. 
If one assumes that the LLs acquire a finite broadening then
all LLs would appear as doubly degenerate except the lowest one in, e.g. an STM measurement. 
In contrast, the LLs are in Figure \ref{fig:LL-gDFTvGW}(b) are more evenly spaced and may 
appear as non-degenerate even if they are broadened. 

Secondly, in some cases  also the sign of $g_{eff}$ changes depending on which parameter set is used. 
For $g_{eff}>0$ the LLs in the $K$ valley have higher energy than the LLs in the $-K$ valley, 
while for negative $g_{eff}$ the opposite is true.  
We note that in Reference\cite{macneill}  
Eqs~\eref{eq:single-band-LL-VB}-\eref{eq:single-band-LL-CB} were used 
            to understand the VDB 
            in the excitonic transitions in MoSe$_2$. 
            The \emph{exciton valley g-factor}  $g_{vl,exc}$ was obtained 
            by considering the energy difference between 
            the lowermost LL in the CB and the uppermost LL in the VB in each valley:
            \begin{eqnarray}
             g_{ex,vl}\mu_B B_z = (E_{n=0,{\rm cb}}^{\tau=1, \downarrow}-E_{n=0,{\rm vb}}^{\tau=1, \downarrow})-
             (E_{n=0,{\rm cb}}^{\tau=-1, \uparrow}-E_{n=0,{\rm vb}}^{\tau=-1,\uparrow}).
            \end{eqnarray}            
             Using Eqs.~\eref{eq:effmass-VB}-\eref{eq:g-CB}, one can easily show  
             that in this approximation the exciton valley g-factor is
             independent of the band gap and it can be expressed  in terms of the effective masses in 
             the CB and VB \cite{macneill,urbaszek2015b}: 
            \begin{eqnarray}
             g_{ex,vl}=4-2\left(\frac{m_e}{m_{cb}^{s}}-\frac{m_e}{|m_{vb}^{s}|}\right)
            \end{eqnarray}  
            Therefore, albeit the effective g-factors  in the CB and VB separately are affected by 
            uncertainties, the exciton g-factor, in principle, can be calculated more precisely 
            so long the effective masses are captured accurately by DFT calculations. 
            The comparison of DFT results and ARPES measurements \cite{kp-review} suggest 
            that the DFT effective masses 
            in the VB  match the experimental results quite well. 
            At the moment, however, it is unclear how accurate are 
            the DFT effective masses in the CB.

Finally, we make the following brief comments.
\begin{itemize}
 \item[i)]  {In the gapped-graphene approximation, i.e.,}  if one neglects the 
            free electron term and the terms $\sim \alpha_{s},\beta_{s}$ 
            in Eqs.~\eref{eq:effmass-VB}-\eref{eq:effmass-CB}  and in \eref{eq:g-VB}-\eref{eq:g-CB} 
            then the lowest LL in the CB and the highest one in the VB will be non-degenerate, 
            but for all other LLs the valley degeneracy would  not be lifted \cite{niu2013b} due to a 
            cancellation effect between the first and last terms in Eqs.\eref{eq:single-band-LL-VB} 
            and \eref{eq:single-band-LL-CB}.
            
\item[ii)]  By measuring the valley g-factors and the effective masses one can deduce the 
            \emph{Diracness} of the spectrum \cite{goerbig}, i.e., the relative importance 
            of the off-diagonal and diagonal terms in 
            $H^{\tau,s}_{D}$ \eref{eq:2dDirac} and $H^{\tau, s}_{\rm as}$ \eref{eq:kp-asym}, respectively. 
            
\end{itemize}


\section{Shubnikov-de Haas oscillations of longitudinal conductivity}
\label{sec:SdH-osc}

As we will show, the results of the Section \ref{subsec:LL-approx} provide 
a convenient starting point for 
the calculation of the SdH oscillations of the  magnetoconductance. 
Our main motivation to consider this problem comes from the 
recent  experimental observation of SdH oscillations in monolayer \cite{hone-SdH} and 
few-layer \cite{hone-SdH,ningwang-SdH} samples.
Regarding previous theoretical works on magnetotransport in TMDCs,  quantum corrections to the 
low-field magneto-conductance were studied in References \cite{shen,OchoaFalko}. 
A different approach, namely, the Adams-Holstein cyclotron-orbit migration theory \cite{holstein},  
was used in Reference~\cite{gzhou} to calculate the longitudinal magnetoconductance $\sigma_{xx}$. 
This theory is applicable if the cyclotron frequency is much larger than the average 
scattering rate $1/\bar{\tau}_{sc}$. 
By using the effective mass  obtained from DFT calculations \cite{kp-review} and 
taking the measured values of the zero field electron mobility 
$\mu_e=\frac{e^2 n_e \bar{\tau}_{sc}}{m_{cb}}$ and the electron density $n_e$ 
given in Reference~\cite{hone-SdH} for monolayer MoS$_2$,  a 
rough estimate for $\bar{\tau}_{sc}$  can be obtained. 
This shows that for  magnetic fields $B\lesssim 15\,$T   the  samples are in the limit of 
$\omega_{cb}\bar{\tau}_{sc} \lesssim 1$ and therefore 
the Adams-Holstein approach cannot be used to describe  $\sigma_{xx}$.  Therefore 
we will  extend  the approach of Ando \cite{ando} to calculate  $\sigma_{xx}$  
in monolayer  TMDCs because it can  
offer a more direct comparison to existing experimental results. 

\begin{figure}[ht]
\begin{center}
 \includegraphics[scale=0.5]{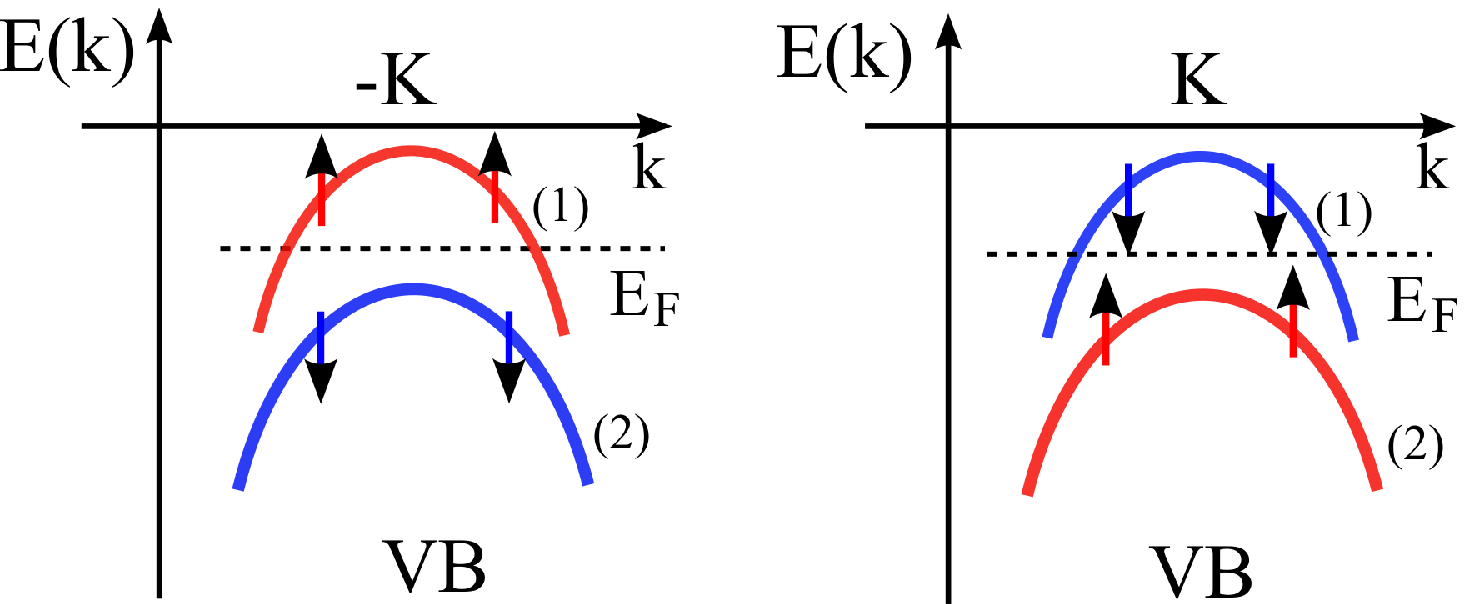}\,\,\,\,\,\,
 \includegraphics[scale=0.5]{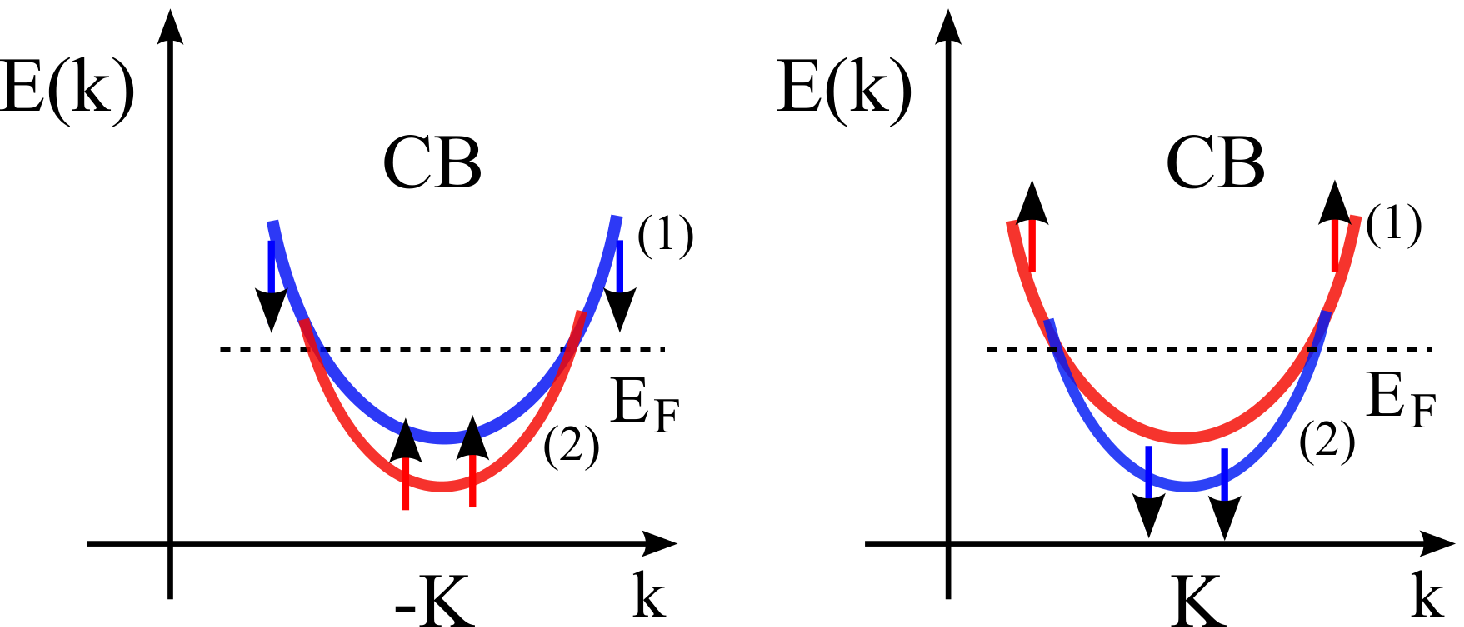}
 \end{center}
 \caption{Schematics of the dispersion in the VB and in the CB around the 
  $K$ and $-K$ points of the band structure. The spin-split bands 
          are denoted by red and blue lines, different colours indicate different 
          spin-polarization. The arrows show the spin-polarization for MoS$_2$. 
         For typical values of  doping, the Fermi-level $E_F$ (denoted by a dashed line) 
         would intersect only the upper spin-split band in the VB or both 
         spin-split bands in the CB. The  index 
         $(1)$ and $(2)$ denote the upper and lower spin-split band.} 
 \label{fig:doping}
\end{figure}

Before presenting the  detailed theory of SdH oscillations we  qualitatively discuss the 
role of the doping and the assumptions that we will use. 
The most likely  scenarios in the VB and the CB are shown in 
Figure~\ref{fig:doping} (a) and (b), respectively.  
Considering first the CB, for electron densities $n_e\sim 10^{13}/{\rm cm^2}$  
measured  in Reference \cite{hone-SdH} 
both the upper and lower spin-split bands would be occupied. 
In contrast, due to the much larger spin-splitting, for hole doped samples  $E_F$ 
would typically intersect only the upper spin-split VB. 
Such a situation may  also occur  for n-doped samples in 
those monolayer TMDCs where the spin-splitting in the CB is much larger 
than in MoS$_2$, e.g., in MoTe$_2$ or WSe$_2$. 
For strong doping other extrema in the VB and CB, such as  the $\Gamma$ and $Q$ points 
may also play a role, this will be briefly discussed at the end of this section.

We will have two main assumptions in the following. The first one is
that one can neglect  inter-valley scattering and also intra-valley scattering between 
the spin-split bands. 
Clearly, this is a simplified model whose validity needs to be checked against experiments. 
One can argue that in the VB  (see Figure \ref{fig:doping}(a)) in the  absence of 
magnetic impurities the inter-valley scattering should be strongly suppressed because  
it would also require a simultaneous spin-flip. 
A recent scanning-tunneling experiment in monolayer WSe$_2$ \cite{leroy2015} 
indeed seems to show a strong  supression of inter-valley scattering. 
In the CB, for the case shown in Figure \ref{fig:doping}(b), the inter-valley scattering is not 
forbidden by spin selection rules. Even if $E_F$ was smaller, such that only one of the 
spin-split bands  is populated in a given valley, the inter-valley scattering would not 
be completely suppressed because the bands are broadened by disorder which can be comparable
to the spin-splitting $2\Delta_{cb}$ ($2\Delta_{cb}=3\,$meV 
for MoS$_2$ and $20-30\,$meV for other monolayer TMDCs.) 
On the other hand,  the intra-valley scattering between the spin-split bands in the CB  
should be absent due to the specific form of the intrinsic SOC, see Eq.~\eref{eq:H_so-K}. 
{We note that strictly speaking  any type of perturbation which breaks the 
 mirror symmetry of the lattice, such as a substrate or certain 
 type of point defects (e.g., sulphur vacancies) would (locally)  
 lead to a Rashba type SOC and hence induce intra-valley coupling between the spin-split bands.
It is not known how effective is this mechanism, in the present study we neglect it.}
The second assumption is that we only consider the effect of short range
scatterers. This assumption is widely used in the interpretation of SdH oscillations as it 
facilitates to obtain  analytical results \cite{ando}. 
We note that according to References~\cite{hone-SdH,goki-eda}, 
some evidence for the presence of short range scatterers in monolayer MoS$_2$ 
has indeed been recently found. 
{ While short-range scatterers can, in general, 
cause inter-valley scattering, on the merit of its simplicity  
as a minimal model  we only take 
into account intra-valley intra-band scattering. }

Using these assumptions it is straightforward to extend the theory of Ando~\cite{ando}  
to the SdH oscillations  of monolayer TMDCs. Namely, as it has been 
shown in Section \ref{subsec:LL-approx}, for not too large magnetic fields 
the LLs in a given band  can be described by a  formula which is the same as  
for a simple parabolic band 
except that it contains a term which describes a linear-in-magnetic field valley-splitting.  
Then, because of  the assumption that one can neglect inter-valley and intra-valley 
inter-band scattering, the 
total conductance will be  the sum of the conductances of  individual bands with valley and spin 
indices $\tau,s$.  This simple model allows us to focus on the effects of intrinsic SOC and 
valley splitting on the SdH oscillations, which is our main interest here. 

Following Reference \cite{ando}, we treat impurity scattering  in the self-consistent Born 
approximation (SCBA) and use the Kubo-formalism to calculate the longitudinal conductivity 
$\sigma_{xx}$ (for a recent discussion see, e.g., \cite{raichev-book,flensberg-book}). 
Assuming  a random  disorder  potential $V(\mathbf{r})$ with short range correlations
 $\langle V(\mathbf{r}) V(\mathbf{r}')\rangle=\lambda_{sc}\delta(\mathbf{r}-\mathbf{r}')$, the 
self-energy $\Sigma^{\tau,s}_{R}=\Sigma^{\tau,s}_r+ i \Sigma^{\tau,s}_i$ in a given band ($\tau,s$) 
does not depend on the LL index $n$. 
It is given by the implicit equation
\begin{eqnarray}
 \Sigma^{\tau,s}_r+ i \Sigma^{\tau,s}_i = \frac{\lambda_{sc}}{2 \pi l_B^2} 
 \sum_{n=0}^{\infty} \frac{1}{E-E_{n}^{\tau,s}-(\Sigma^{\tau,s}_r+ i \Sigma^{\tau,s}_i)}
\label{eq:Sigma-complx} 
\end{eqnarray}
where $E_{n}^{\tau,s}$ is given by Eqs~\eref{eq:single-band-LL-VB}-\eref{eq:single-band-LL-CB}. 
The  term $\lambda_{sc}/2 \pi l_B^2$ on the right-hand side of Eq.\eref{eq:Sigma-complx}   
can be rewritten as 
$
\frac{\lambda_{sc}}{2 \pi l_B^2}=\frac{1}{2\pi} \hbar\omega_{c}^{(i)}\frac{\hbar}{\tau_{sc}^{(i)}}
$
where $1/\tau_{sc}^{(i)}=\lambda_{sc} m^{(i)}/\hbar^3$ is the scattering rate 
calculated in the Born-approximation
in zero magnetic field. As in Section \ref{subsec:LL-approx}, 
the upper index $i=1 (2)$ refers to the higher(lower)-in-energy
spin-split band in a given valley (see also Figure~\ref{fig:doping}). 
Using the Kubo-formalism the conductivity coming from a single valley and band 
$\sigma_{xx}^{\tau,s}$ is calculated as 
\begin{eqnarray}
 \sigma_{xx}^{\tau,s}=\frac{e^2}{\pi^2 \hbar}\int \mathnormal{d}E
 \left(-\frac{\partial f(E)}{\partial E}\right)
 \sigma_{xx}^{\tau,s}(E)
 \label{eq:sigmaxx}
 \end{eqnarray}
 where $f(E)$ is the Fermi function and 
\begin{eqnarray} 
\fl
 \sigma_{xx}^{\tau,s}(E)=(\hbar\omega_{c}^{(i)})^2\sum_{n=0}^{\infty} (n+1) 
 {\rm Re}[&G^{\tau,s}_{A}(n,E)G^{\tau,s}_{R}(n+1,E)-
                                    &G^{\tau,s}_{A}(n,E)G^{\tau,s}_{A}(n+1,E)].
\label{eq:sigmaxx-Greens}                                    
\end{eqnarray}
Here $G^{\tau,s}_{R}(n,E)$ and $G^{\tau,s}_{A}(n,E)$ are the retarded and 
advanced Greens-functions, respectively.  Vertex corrections are neglected in this approximation.   
Since we neglect inter-valley and intra-valley inter-band scattering, 
the disorder-averaged Greens-function
$ G^{\tau,s}_{R,A}(n,E)=[E-E_{n}^{\tau,s}-\Sigma^{\tau,s}_{R,A}]^{-1}$
is diagonal in  the indices $\tau,s$ and  in the LL  representation it is 
also diagonal in the LL index $n$.
The total conductivity is then given by 
 $\sigma_{xx}=\sum_{\tau,s}\sigma_{xx}^{\tau,s}$
where the summation runs over occupied subbands for a given total electron (hole) 
density $n_e$ ($n_h$). 
In general, one has to  determine $ \Sigma^{\tau,s}_r+ i \Sigma^{\tau,s}_i $ 
by soving  Eq.~\eref{eq:Sigma-complx} numerically. The Greens-functions $G^{\tau,s}_{R,A}$ 
can  be then calculated and $\sigma_{xx}^{\tau,s}$ follows from Eq.~\eref{eq:sigmaxx-Greens}. 
It can be seen from  Eq.~\eref{eq:sigmaxx} that at zero temperature 
$\Sigma^{\tau, s}(E)$ and $\sigma_{xx}^{\tau,s}(E)$
has to be evaluated at $E=E_F$. In the semiclassical limit, 
when there are many occupied LLs below  $E_F$, i.e., $\hbar\omega^{(i)}_{c}\ll E_F$, one can derive 
an analytical result for   $\sigma_{xx}^{\tau,s}$, see References \cite{ando,raichev-book} for 
the details of this calculation. Here we only give the final form of  $\sigma_{xx}$  
and compare it to the results of numerical calculations.

As mentioned above, the situation depicted in Figure~\ref{fig:doping}(a), i.e.,  when there is only 
one occupied subband in each of the valleys is probably most relevant for  p-doped samples.  
One finds that in this case the longitudinal conductance is 
\begin{eqnarray}
\fl
\sigma_{xx}/\sigma_0=\frac{2}{1+(\omega_{\rm vb}^{(1)}\tau_{sc}^{(1)})^2}
 \left[1-\frac{4 (\omega_{\rm vb}^{(1)}\tau_{sc}^{(1)})^2}{1+(\omega_{\rm vb}^{(1)}\tau_{sc}^{(1)})^2}
 e^{-\frac{\pi}{(\omega_{\rm vb}^{(1)}\tau_{sc}^{(1)})}}
 \cos\left(\frac{2 \pi E_F}{\hbar\omega_{\rm vb}^{(1)}}\right)\mathcal{A}_1\,\mathcal{B}\nonumber\right.\\
 +\left. \frac{g_{eff}^{(i)}}{2}\frac{\mu_B B_z}{E_F} 
 \frac{4 (\omega_{\rm vb}^{(1)}\tau_{sc}^{(1)})^2}{1+(\omega_{\rm vb}^{(1)}\tau_{sc}^{(1)})^2}
 e^{-\frac{\pi}{(\omega_{\rm  vb}^{(1)}\tau_{sc}^{(1)})}}
 \sin\left(\frac{2 \pi E_F}{\hbar\omega_{\rm vb}^{(1)}}\right)\mathcal{A}_2\,\mathcal{B}\right].
 \label{eq:sigmaxx-two-valley-one-band}
\end{eqnarray}
Here 
$
\sigma_0=\frac{e^2\tau_{sc}^{(1)}}{2 \pi \hbar^2} E_F=
\frac{e^2\tau_{sc}^{(1)}}{m_{\rm vb}^{(1)}} \frac{n_{h}}{2}
$ 
is the zero field conductivity per single valley and band, $n_{h}$ is the total 
charge density and we assumed  $\Sigma^{\tau,s}_r \ll \Sigma^{\tau,s}_i \ll E_F$. 
The amplitudes $\mathcal{A}_{1,2}$ and $\mathcal{B}_{}$ are given by 
\numparts
\begin{eqnarray}
 \mathcal{A}_{1}=\cos\left(\frac{\pi}{2}g_{eff,{\rm vb}}^{(1)}\frac{m_{\rm vb}^{(1)}}{m_e}\right),
 \hspace{1cm}
 \mathcal{A}_{2}=\sin\left(\frac{\pi}{2}g_{eff,{\rm vb}}^{(1)}\frac{m_{\rm vb}^{(1)}}{m_e}\right);\\
\mathcal{B}_{}=\frac{2 \pi^2 k_B T/\hbar \omega_{\rm vb}^{(1)}} 
{\sinh\left({2 \pi^2 k_B T/\hbar \omega_{\rm vb}^{(1)}}\right)},
\label{eq:amplitude-two-valley-one-band}
\end{eqnarray}
\endnumparts
where $k_B$ is the Boltzmann constant and $T$ is the temperature.
One can see that Eqs.~\eref{eq:sigmaxx-two-valley-one-band}-\eref{eq:amplitude-two-valley-one-band} 
are very similar to the  well known expression derived by Ando \cite{ando} 
for a two-dimensional electron gas (2DEG). 
The valley-splitting, which leads to the appearance of the amplitudes $\mathcal{A}_{1,2}$,  
plays an analogous role to the Zeeman spin-splitting in 2DEG. 
Therefore, under the assumption we made above,  
the uncertainty regarding the value of the effective g-factors affects the amplitude of 
the oscillations but not their phase.  
The term proportional  to $\mu_B B_z/E_F$  in  Eq.~\eref{eq:sigmaxx-two-valley-one-band} is usually  
much smaller than the first term. Thus, it  can be  neglected  in the calculation of 
the total conductance, but may be important if one is interested 
only in the oscillatory part of  $\sigma_{xx}$, see below. 

\begin{figure}[htb]
\begin{center}
 \includegraphics[scale=0.5]{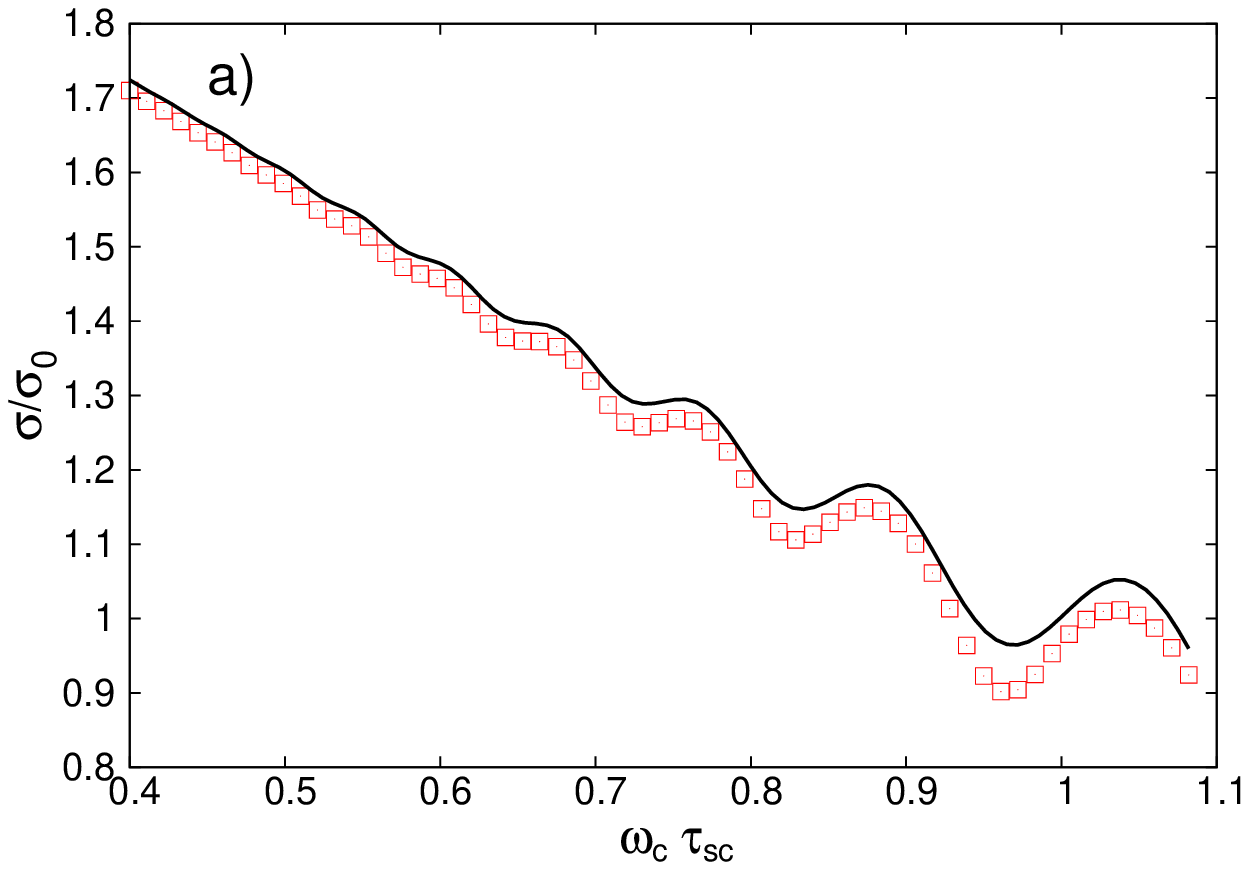}\,\,\,
 \includegraphics[scale=0.5]{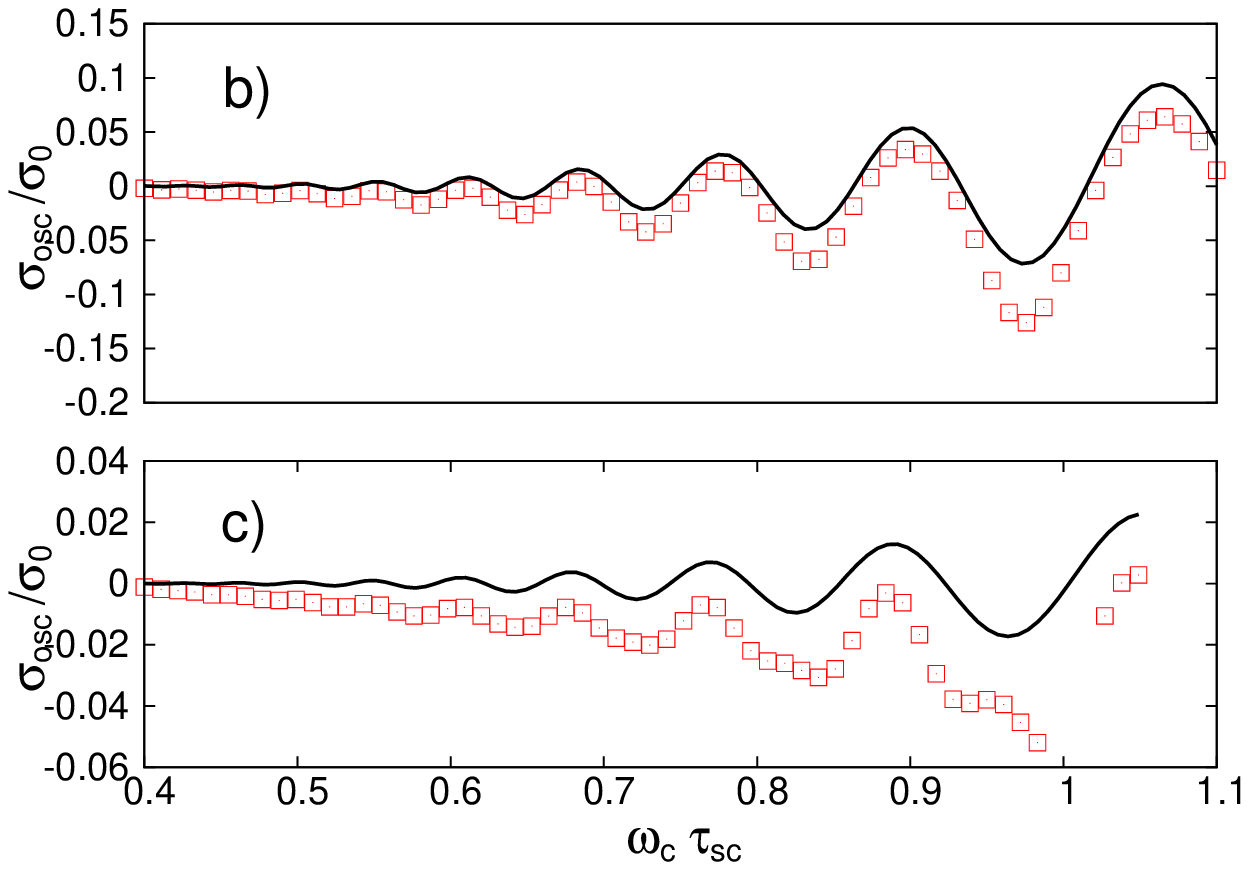}
 \end{center}
 \caption{Comparison of the numerically (symbols) and analytically (solid line) calculated zero 
  temperature conductivity for WSe$_2$ for  the situation depicted in Figure \ref{fig:doping}. 
  a) total conductivity, $g_{eff, {\rm vb}}^{(1)}=0.55$.  
  b) and c) comparison of the oscillatory parts of $\sigma_{xx}$. 
  In b) $g_{eff, {\rm vb}}^{(1)}=0.55$, while in c) $g_{eff, {\rm vb}}^{(1)}=-2.38$ 
  (see Table \ref{tbl:g-factors-WSe2}). The figures correspond to a magnetic field 
   range of about $5.7-15.7\,$T.} 
 \label{fig:conduct-WSe2}
\end{figure}

We emphasize that Eq.~\eref{eq:sigmaxx-two-valley-one-band} is only accurate if 
$\hbar\omega_{\rm vb}^{(1)}\ll E_F$. 
However, in semiconductors, especially at relatively low doping, 
one can reach magnetic field values where the 
cyclotron energy becomes comparable to $E_F$. In this case the numerically calculated 
$\sigma_{xx}$  may differ from Eq.~\eref{eq:sigmaxx-two-valley-one-band}
\footnote{{From a theoretical point of view, 
in strong magnetic fields one should also calculate  vertex correlations to $\sigma_{xx}$, 
but this is not considered here.}}. 
It is known that, e.g., WSe$_2$ can be relatively easily gated into the VB, 
and a decent  Hall mobility 
was recently demonstrated in few-layer samples in Reference \cite{jarillo-herrero2015}.  
As a concrete example  we take the following  values \cite{jarillo-herrero2015}: 
$n_h=-4*10^{12}/cm^2$ and Hall mobility $\mu_H=700 cm^2/Vs$. 
By taking  $m_{vb}^{(1)}=-0.36 m_e$ \cite{kp-review} the Fermi energy is 
$E_{F}\approx-26.6$meV and  using that $\mu_H=e \tau_{sc}^{(1)}/m^{(1)}_{vb}$ 
we obtain $\tau_{sc}^{(1)}=1.4 \times 10^{-13}$s. The amplitude of the 
oscillations should become discernible when 
$\omega_{vb}^{(1)}\tau_{sc}^{(1)}=\mu_H B_z \lesssim 1$, i.e., 
for magnetic 
fields $B_z\gtrsim 10$T, while at $B_z=14.28$T, which corresponds to 
$\omega_{vb}^{(1)}\tau_{sc}^{(1)}\approx 1$, 
there are around six occupied LLs.  
One can expect that for  $B_z\lesssim 15\,$T the LL spectrum is well 
described by Eq\eref{eq:single-band-LL-VB}, however,   
since the number of LLs is relatively low, there might be deviations between the analytically 
and numerically calculated $\sigma_{xx}$. 
In Figure \ref{fig:conduct-WSe2}(a) we  show a comparison between 
the analytical result Eq.~\eref{eq:sigmaxx-two-valley-one-band} and 
the numerically calculated longitudinal conductance at zero temperature. 
The  effective g-factors $g_{eff,{\rm vb}}^{(1)}$ used in these calculations 
are given in Table \ref{tbl:g-factors-WSe2}. 

One can see that for larger magnetic fields the amplitude of the oscillations 
is not captured very precisely by 
Eq.~\eref{eq:sigmaxx-two-valley-one-band} but the overall agreement with the 
numerical results is  good.
Next, in Figures~\ref{fig:conduct-WSe2}(b) and (c) we compare the
oscillatory parts $\sigma_{xx,osc}$ of the 
longitudinal conductivity obtained from numerical calculations 
and from Eq~\eref{eq:sigmaxx-two-valley-one-band} 
using two different $g_{eff, {\rm vb}}$ values. 
In the case of the numerical calculations $\sigma_{xx,osc}$ was obtained by  
subtracting the smooth function $2/(1+(\omega_{\rm vb}^{(1)}\tau_{sc}^{(1)})^{2})$ 
from $\sigma_{xx}$. 
According to Eq.~\eref{eq:sigmaxx-two-valley-one-band}, the valley-splitting of the LLs and the 
different effective g-factors should only affect the amplitude of the oscillations. 
While the amplitude of the oscillations  
indeed depends on $g_{eff, {\rm vb}}^{(1)}$, one can see that the agreement is better for 
$g_{eff, {\rm vb}}^{(1)}=0.55$ than for $g_{eff, {\rm vb}}^{(1)}=-2.38$. 
For the latter case the position of the conductance minimuma start to differ 
for large magnetic field, 
whereas the  maximuma in $\sigma_{xx}$ agree in both figures. These calculations illustrate that 
Eq.~\eref{eq:sigmaxx-two-valley-one-band} may not agree with the numerical results 
when there are only a few LLs below $E_F$.

We now turn to the case shown in Figure \ref{fig:doping}(b) when both spin-split subbands are 
populated. The total conductance is given by the sum of the conductances coming 
from the two spin-split subbands : $\sigma_{xx}=\sigma_{xx}^{(1)}+\sigma_{xx}^{(2)}$. 
Since the effective masses in the spin-split bands are, in general, different, the 
associated scattering times 
$\tau_{sc}^{(1)}$ and $\tau_{sc}^{(2)}$ calculated in the Born-approximation are also different.
We define 
$\tilde{\tau}_{sc}=\tau_{sc}^{(1)}+\tau_{sc}^{(2)}$ and 
$\tilde{\sigma}_0=\frac{e^2 \tilde{\tau}_{sc}^{}}{2 \pi \hbar^2} E_F$, 
and obtain for the magnetoconductance  
\begin{eqnarray}
\fl
\sigma_{xx}/\tilde{\sigma}_0=
  2\mathcal{C}^{(2)}\frac{1}{1+(\omega_{\rm cb}^{(2)}\tau_{sc}^{(2)})^2}
\left[1-\frac{4 (\omega_{\rm cb}^{(2)}\tau_{sc}^{(2)})^2}{1+(\omega_{\rm cb}^{(2)}\tau_{sc}^{(2)})^2}
  e^{-\frac{\pi}{(\omega_{\rm cb}^{(2)}\tau_{sc}^{(2)})}}
 \cos\left(\frac{2 \pi E_F}{\hbar\omega_{\rm cb}^{(2)}}\right)
 \mathcal{A}_{1}^{(2)}\,\mathcal{B}^{(2)}\right]\nonumber\\
 \fl
 + 2 \mathcal{C}^{(1)}\frac{1}{1+(\omega_{\rm cb}^{(1)}\tau_{sc}^{(1)})^2}
 \left[1-\frac{4 (\omega_{\rm cb}^{(1)}\tau_{sc}^{(1)})^2}{1+(\omega_{\rm cb}^{(1)}\tau_{sc}^{(1)})^2} e^{-\frac{\pi}{(\omega_{\rm cb}^{(1)}\tau_{sc}^{(1)})}}
 \cos\left(\frac{2 \pi (E_F-2\Delta_{cb})}{\hbar\omega_{\rm cb}^{(1)}}\right)
 \mathcal{A}_{1}^{(1)}\,\mathcal{B}^{(1)}\right]
 \label{eq:sigmaxx-two-valley-two-band}
\end{eqnarray}
Here 
\numparts
\begin{eqnarray}
 \mathcal{C}^{(1)}=\left(1-\frac{2\Delta_{cb}}{E_F}\right)\frac{\tau_{sc}^{(1)}}{\tilde{\tau}_{sc}},
 \hspace{1cm}
 \mathcal{C}^{(2)}=\frac{\tau_{sc}^{(2)}}{\tilde{\tau}_{sc}}, \label{eq:C1}\\
 \mathcal{A}_{1}^{(i)}=\cos\left(\frac{\pi}{2}g_{eff}^{(i)}\frac{m_{{\rm cb}}^{(i)}}{m_e}\right), 
 \label{eq:A1}\\
\mathcal{B}^{(i)}=\frac{2 \pi^2 k_B T/\hbar \omega_{\rm cb}^{(i)}}
{\sinh\left({2 \pi^2 k_B T/\hbar \omega_{\rm cb}^{(i)}}\right)}.
\end{eqnarray}
\endnumparts
In Eq.~\eref{eq:sigmaxx-two-valley-two-band} we have neglected terms which are $\sim\mu_B B_z/E_F$.
The result shown in Eq.~\eref{eq:sigmaxx-two-valley-two-band} 
is  similar to the multiple subband occupation problem in 2DEG \cite{raikh,willander,raichev}. 
The valley splitting  affects the amplitude of the oscillations (see Eq.~\eref{eq:A1}), 
whereas the intrinsic SOC can influence the amplitude of the oscillations [see Eq.~\eref{eq:C1}]
and  it also leads to a phase difference [Eq.~\eref{eq:sigmaxx-two-valley-two-band}] 
between the oscillations coming from the two spin-split subbands. 

The situation depicted in Figure \ref{fig:doping}(b) is easily attained, e.g., in the CB of 
monolayer  MoS$_2$, where DFT calculations predict  that the  spin-splitting is 
$2\Delta_{cb}=3$\,meV and therefore both subbands can be populated for relatively low
densities. 
Our choice of the parameters for  the numerical calculations shown below is motivated by the 
recent experiment of  Cui \emph{et al.} \cite{hone-SdH}, where 
SdH oscillations in mono- and few layer MoS$_2$ samples have been measured.
We use $n_e=10^{13}/cm^2$ and mobility $\mu_H=1000 {\rm cm^2/V s}$.
The effective masses are chosen as  $m_{cb}^{(1)}=0.46 m_e$, $m_{cb}^{(2)}=0.43 m_e$
and the spin-splitting in the CB is $2\Delta_{cb}=3$\,meV \cite{kp-review}. 
Using these parameters we find $E_F=28.43$\,meV.   
Since  the effective masses are rather similar, 
the scattering times calculated from $\mu_H$ are close to each other: 
$\tau_{sc}^{(1)}\approx \tau_{sc}^{(2)} \approx 2.6 \times 10^{-13}$s, 
i.e., they are almost twice as long as in the case of WSe$_2$. 
The oscillations in $\sigma_{xx}$ should become discernible 
for $B_z\gtrsim 7\,$T, and at $B_z=10$T  there are ten LLs in both the lower and the 
upper spin-split CB in each valley.  We will focus on the oscillatory part 
$
\sigma_{xx,osc}=\sigma_{xx,osc}^{(1)}+\sigma_{xx,osc}^{(2)}
$ 
of the conductance, since this contains information about  the spin and valley splittings. 
As in the previous example, we first calculate $\sigma_{xx}^{(i)}$ 
numerically using Eqs.~\eref{eq:Sigma-complx}-\eref{eq:sigmaxx-Greens} and obtain 
$\sigma_{xx,osc}^{(i)}$  by subtracting the smooth
function $2\mathcal{C}^{(i)}/[1+(\omega^{(i)}_{c}\tau_{sc}^{(i)})^2]$. We than compare these results 
to the oscillations that can be obtained from Eq.~\eref{eq:sigmaxx-two-valley-two-band}. 

\begin{figure}[ht]
\begin{center}
 \includegraphics[scale=0.5]{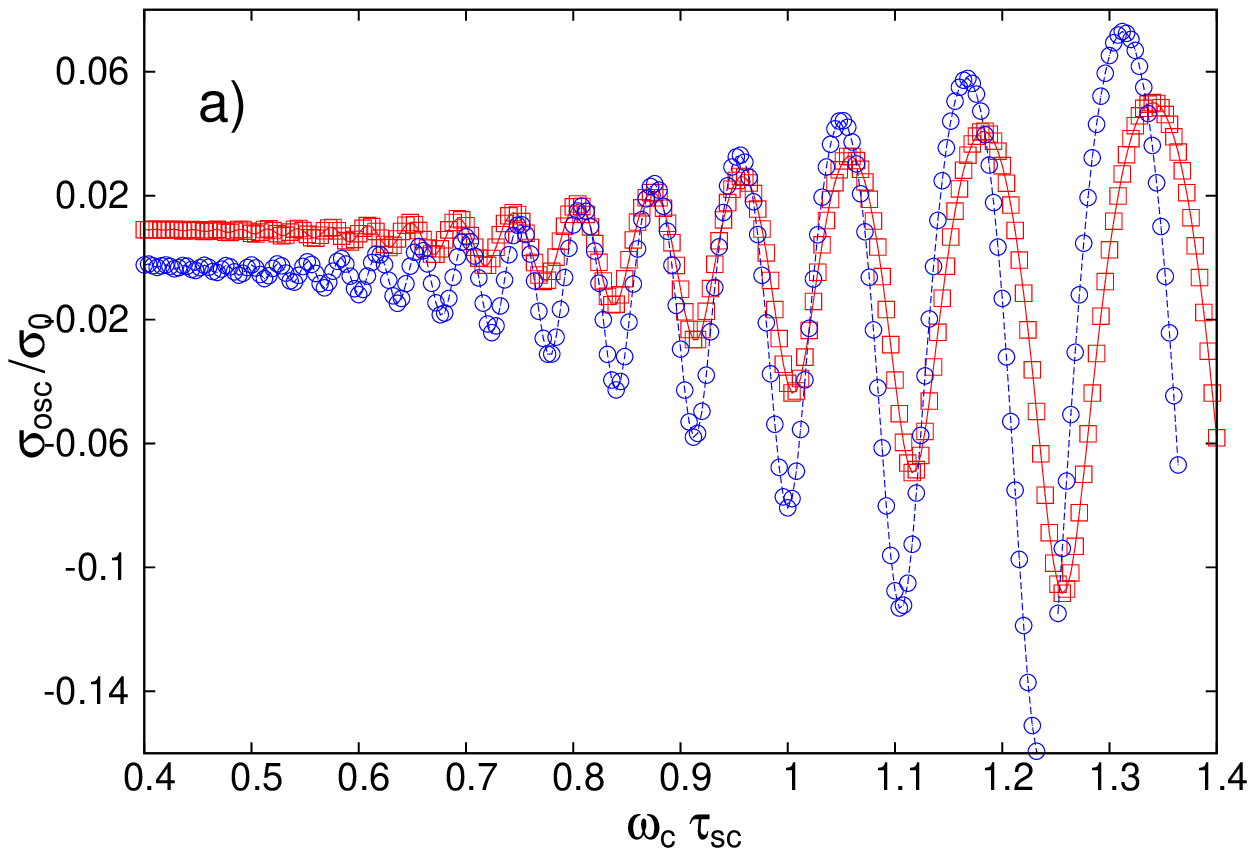}\,\,\,\,
 \includegraphics[scale=0.5]{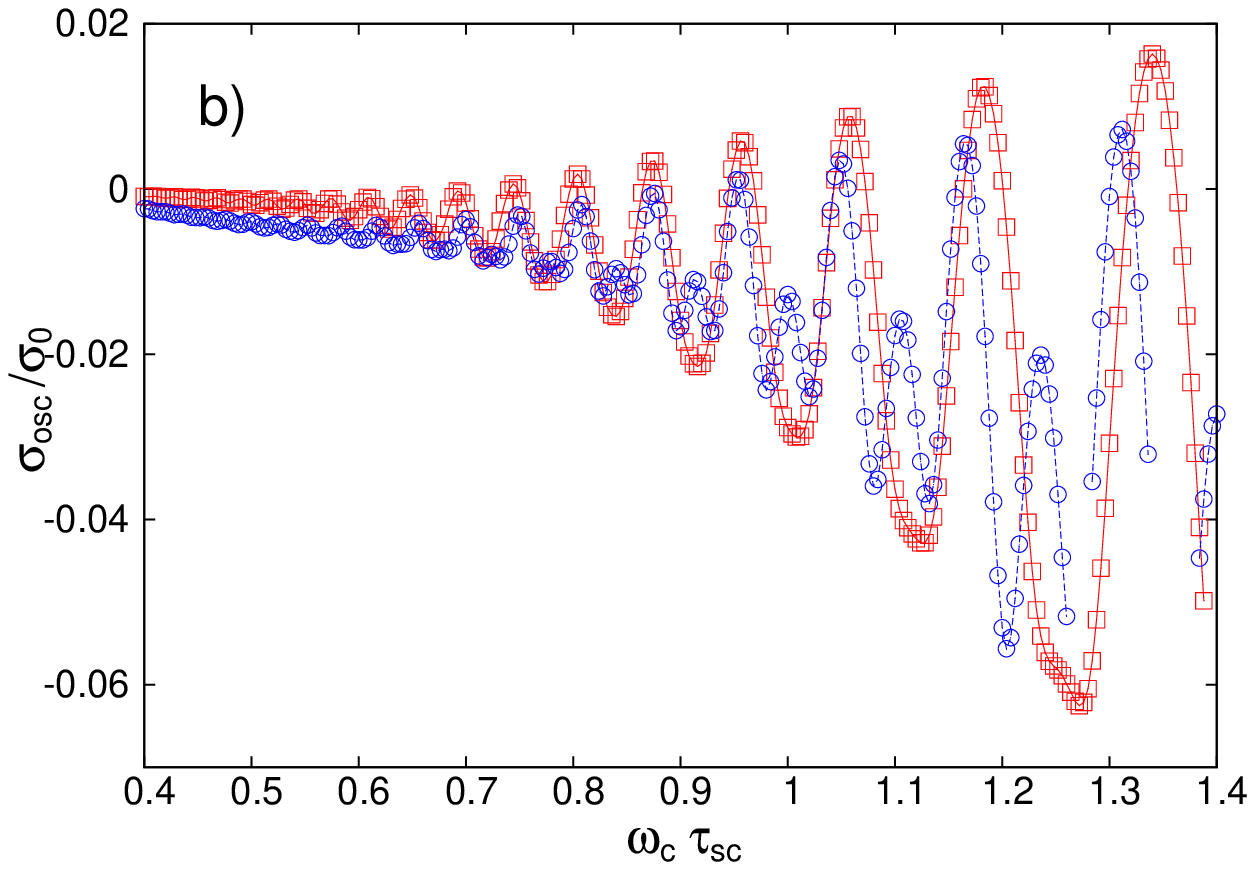}\\
 \includegraphics[scale=0.5]{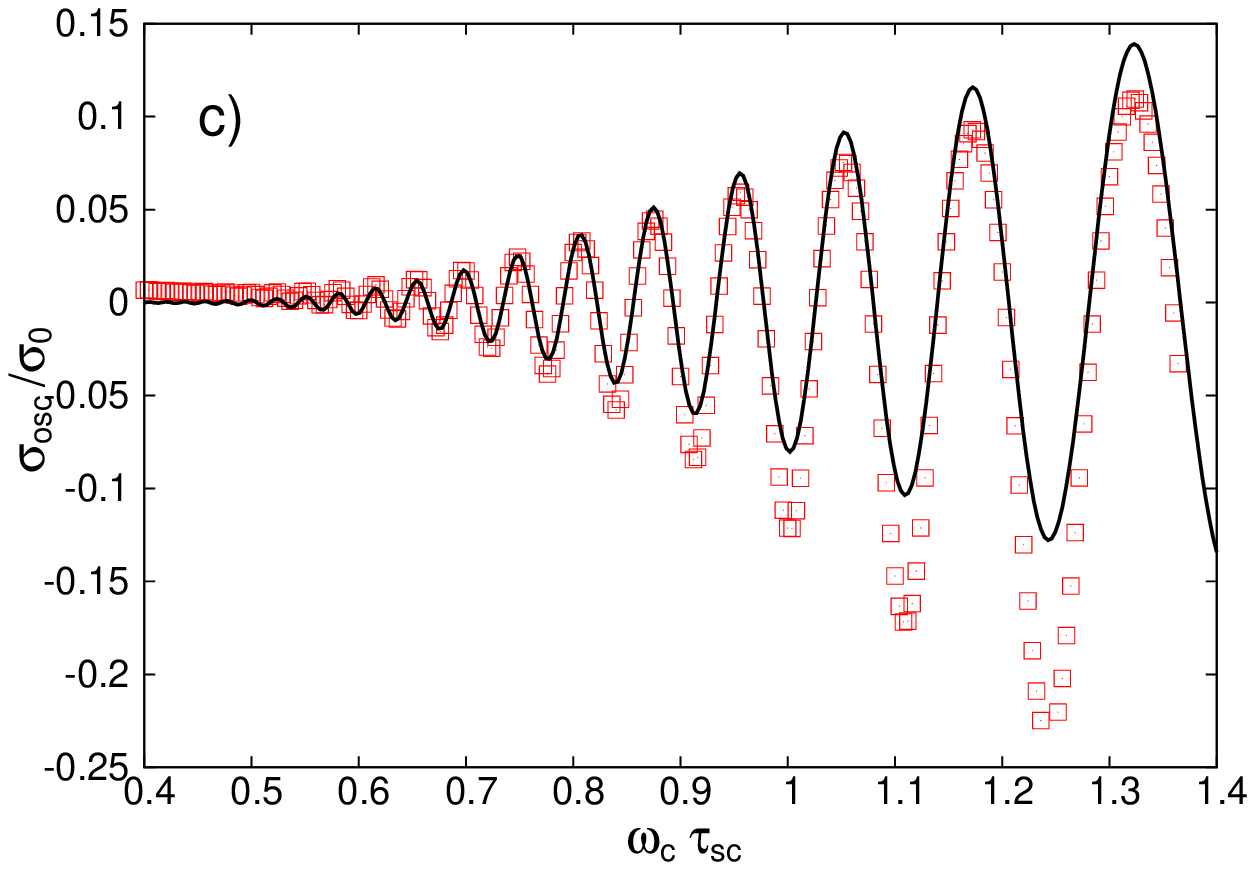}\,\,\,\,
 \includegraphics[scale=0.5]{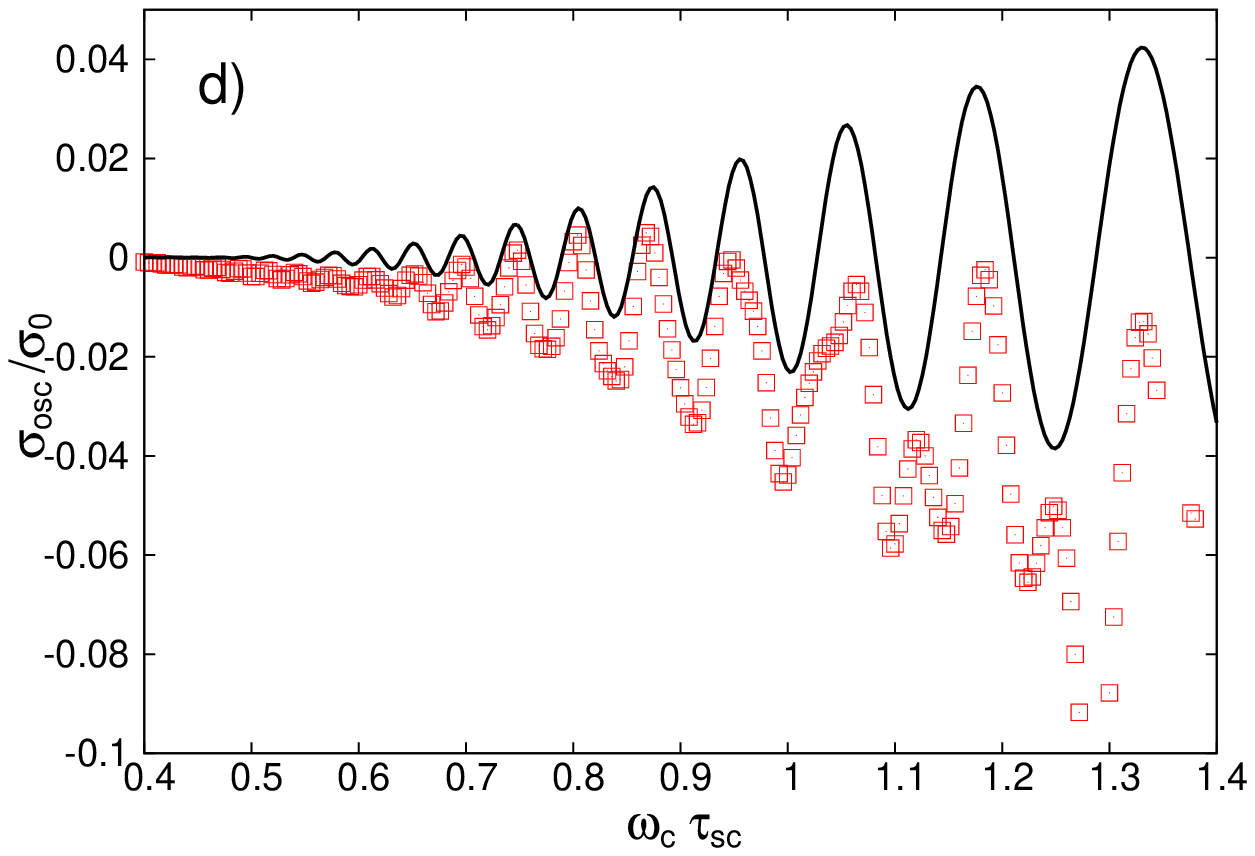}
 \end{center}
 \caption{Oscillations of $\sigma_{xx}$ in n-doped MoS$_2$. a) numerically calculated 
  $\sigma_{osc}^{(1)}$ (red squares) and $\sigma_{osc}^{(2)}$ (blue circles) using  
  $g_{eff, {\rm cb}}^{(1)}=-0.05$ and $g_{eff, {\rm cb}}^{(2)}=-4.11$. 
  b)  numerically calculated $\sigma_{osc}^{(1)}$ 
  (red squares) and $\sigma_{osc}^{(2)}$ (blue circles) using  $g_{eff, {\rm cb}}^{(1)}=1.44$ and 
 $g_{eff, {\rm cb}}^{(2)}=-2.55$. 
 c) The total oscillatory conductance $\sigma_{osc}=\sigma_{osc}^{(1)}+\sigma_{osc}^{(2)}$ 
    corresponding to a) (red squares) and the analytical result calculated from 
    Eq.~\eref{eq:sigmaxx-two-valley-two-band} (solid line).
    d) the same as in c) but corresponding to b).The figures correspond to a magnetic 
    field range of about $4-14\,$T.} 
 \label{fig:oscillations-MoS2}
\end{figure}

In Figures \ref{fig:oscillations-MoS2}(a) and (b) we show the numerically calculated  
$\sigma_{xx}^{(1)}$ and $\sigma_{xx}^{(2)}$ for the two different sets of g-factor  values 
given in Table \ref{tbl:g-factors-MoS2} as a function of $\bar{\omega}_{c}\bar{\tau}_{sc}$
which was introduced as a dimensionless scale of the magnetic field.
Here $\bar{\omega_{c}}=\frac{e B_z}{\bar{m}_{cb}}$ with $\bar{m}=\sqrt{m_{cb}^{(1)}m_{cb}^{(2)}}$ and 
$\bar{\tau}_{sc}=\sqrt{\tau_{sc}^{(1)}\tau_{sc}^{(2)}}$. 
All calculations are at zero temperature. 
One can observe that due to the CB spin splitting $2\Delta_{cb}$ the oscillations of 
$\sigma_{xx}^{(1)}$ and $\sigma_{xx}^{(2)}$  will not be in-phase for larger  magnetic field. 
This effect is expected to be 
even more important for TMDCs having larger $2\Delta_{cb}$ than MoS$_2$ and 
leads to  more complex oscillatory features in the total conductance $\sigma_{xx}^{}$ 
than in the previous example of p-doped WSe$_2$ where only one band in each valley 
contributed to the conductance. One can also observe that in Figure \ref{fig:oscillations-MoS2}(b) 
additional peaks with smaller amplitude appear in $\sigma_{xx,osc}^{(2)}$ for larger magnetic fields, 
while there are no such peaks in  $\sigma_{xx,osc}^{(2)}$ in Figure \ref{fig:oscillations-MoS2}(a). 
The origin of this behaviour can be traced back to the different valley-splitting patterns 
shown in  Figure \ref{fig:LL-gDFTvGW}. 
The valley splitting of the LLs  in Figure \ref{fig:LL-gDFTvGW}(a) is small 
(except for the lowest LL), while 
in Figure \ref{fig:LL-gDFTvGW}(b) all LLs belonging to different valleys are 
well-separated for larger fields 
and this leads to the appearance of the additional, smaller amplitude 
peaks in $\sigma_{xx,osc}^{(2)}$ 
in Figure \ref{fig:oscillations-MoS2}(b). 
The comparison between the numerically calculated total oscillatory part 
$\sigma_{xx,osc}^{}=\sigma_{xx,osc}^{(1)}+\sigma_{xx,osc}^{(2)}$ 
and the corresponding analytical result given in 
Eq.~\eref{eq:sigmaxx-two-valley-two-band} is shown in Figures \ref{fig:oscillations-MoS2}(c) and (d). 
The agreement between the two approaches is qualitatively good for 
$\bar{\omega}_{c}\bar{\tau}_{sc}^{}\lesssim 1$. 
However, for larger magnetic fields  where $\bar{\omega}_{c}\bar{\tau}_{sc}^{}\gtrsim 1$ 
the amplitude of the oscillations start to differ. 
In this regime the  oscillatory behaviour in  $\sigma_{xx,osc}^{}$ can be quite complex, 
influenced by both the valley splitting  and also by the intrinsic SOC splitting of the bands.

We have tried to analyze the experimental results by Cui \emph{et al.} \cite{hone-SdH} using 
the theoretical approach outlined above. 
To this end we have first calculated $\sigma_{xx,exp}(B_z)$ by  inverting the 
experimentally obtained resistance matrix and normalized it by the zero-field conductance 
$\sigma_{xx,exp}(0)$.  
To simplify the ananlysis, we assumed that the effective masses are the 
same in the two spin-split CB: $m_{\rm cb}^{(1)}=m_{\rm cb}^{(2)}=0.43 m_e$, 
and hence $\tau_{sc}^{(1)}=\tau_{sc}^{(2)}$. 
We then fitted  $\sigma_{xx,exp}(B_z)/\sigma_{xx}(0)$ by the function $f_{0}(B_z)=C+A/(1+ (\mu_q B_z)^2)$,
where the amplitudes $A$, $C$ and the quantum mobility $\mu_{q}$ are fit parameters. 
This function, according to Eq.~\eref{eq:sigmaxx-two-valley-two-band}, should give the smooth
part of the conductance. The fit was performed in the magnetic field range
$[4{\rm T}-15{\rm T}]$: for smaller fields the weak-localization corrections might
be important which are not considered in this work, while in larger magnetic field   
the semiclassical approximation may not be accurate. 
We have found that $\sigma_{xx,exp}$ can be approximated quite  
well by $f_{0}(B_z)$. The most important parameter that can be extracted from 
the fit is the quantum scattering time $\tau_{sc,q}$, which is obtained from 
$\tau_{sc,q}=\frac{m_{\rm cb} \mu_q}{e}$.
We find that it is roughly $3.5$ times shorter than the transport scattering time $\tau_{sc,tr}$
that follows  from the measured Hall mobility $\mu_H=1000\,{cm}^{2}/{\rm V s}$. 
The ratio $\tau_{sc,tr}/\tau_{sc,q}$ depends to some extend on the fitting range that is used, but  
typically it is  $\tau_{sc,tr}/\tau_{sc,q}> 2$.  
This difference may be explained by the fact that small-angle scattering is unimportant for 
 $\tau_{sc,tr}$ but it can affect $\tau_{sc,q}$. 
We note that  Cui \emph{et. al.} \cite{hone-SdH} has also found that the 
$\tau_{sc,tr}$ is larger than $\tau_{sc,q}$, but they have used the amplitude of the 
longitudinal resistance oscillations in the magnetic field range $10-25$T to extract $\tau_{sc,q}$ 
and obtained $\tau_{sc,tr}/\tau_{sc,q}\approx 1.5$. 
The significantly shorter $\tau_{sc,q}$ makes it difficult to analyze the magnetic oscillations 
in a quantitative way using Eq.~\eref{eq:sigmaxx-two-valley-two-band}.
Namely, it implies that oscillations should be discernible for $B_z\gtrsim 15$T, i.e., for 
magnetic fields where  only a few LLs are occupied and the semiclassical approximation 
may not be accurate. 
\begin{figure}[ht]
 \begin{center}
\includegraphics[scale=0.6]{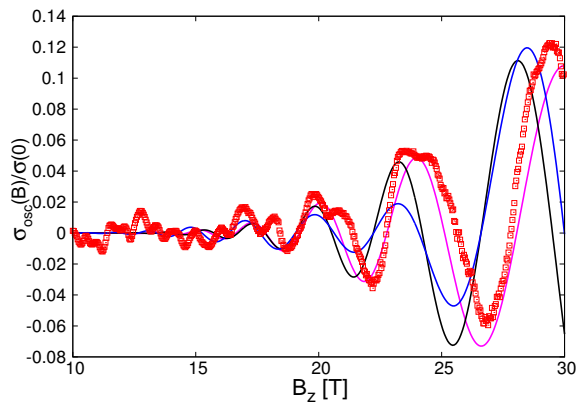}
  \end{center}
  \caption{Comparison of the theoretical results and the measurements of 
           Cui \emph{et al.} \cite{hone-SdH}. The measured conductance 
           oscillations $\sigma_{xx,exp}^{osc}$ (squares) and the fitting of the  
           function  $f_{osc}(B_z)$ (see Eq.~\eref{eq:fit-osc}) 
           using $n_e=1.1*10^{13}/{\rm cm}^2$, $2\Delta_{cb}=3\,$meV (blue) 
           $n_e=1.31*10^{13}/{\rm cm}^2$, $2\Delta_{cb}=3\,$meV (black), 
            and $n_e=1.31*10^{13}/{\rm cm}^2$, $2\Delta_{cb}=5\,$meV (purple).} 
  \label{fig:exp-and-theor}
 \end{figure}
Using $f_{0}(B_z)$ we then extracted the  oscillatory part 
 $\sigma_{xx,osc}(B_z)=\sigma_{xx,exp}(B_z)-f_{0}(B_z)$ of the conductance and fitted 
 it with the function 
 \begin{equation}
 \fl
  f_{osc}(B_z)= -A\frac{4 (\mu_q B_z)^2}{(1+(\mu_q B_z)^2)^2}\exp\left(\frac{-\pi}{\mu_q B_z}\right)
  \left[D_1 \cos\left(\frac{2\pi E_F}{\hbar\omega_{\rm cb}}\right)
        +D_2\cos\left(\frac{E_F-2\Delta_{cb}}{\hbar\omega_{\rm cb}}\right)\right]      
\label{eq:fit-osc}
 \end{equation}
 where $D_{1,2}$ are fitting parameters. As one can see in Figure \ref{fig:exp-and-theor},
the  fit can qualitatively reproduce the meauserements, but the complex oscillations 
 between $15-22$T are not captured. We also note that a somewhat better fit can be obtained 
 if we assume that the charge density is larger than what is deduced from the 
 classical Hall measurements {(see the black line in Figure \ref{fig:exp-and-theor})
  and if we choose the spin-splitting larger than the value obtained from DFT 
  calculations (purple line). In all cases we find, however, that 
  the fit parameters $D_{1}$ and $D_{2}$ differ quite significantly in their magnitude,
  which is difficult to interpret in the present theoretical framework. 
  This might indicate that additional scattering channels, such as inter-valley
  scattering, would have to be taken into account for a quantitative theory. }

Finally, we would briefly comment on the relevance of the other valleys in the band structure  for 
the SdH oscillations. Regarding p-doped samples, the $\Gamma$ point might, in principle, be 
important for MoS$_2$.  However, according to DFT calculations \cite{kp-review} and 
ARPES measurements \cite{osgood2013} the effective mass at the $\Gamma$ point is significantly 
larger than at $\pm K$ and therefore  we do not expect that  
states at $\Gamma$ would lead to additional SdH oscillations. Nevertheless, they 
can be important for the level broadening of the sates  at $\pm K$ because 
scattering from $\pm K$ to $\Gamma$ does not require a spin-flip \cite{kp-review}. 
In the case of other  monolayer TMDCs the $\Gamma$  valley is most likely too far away in energy 
from the top of the VB at $\pm K$ to influence the transport for realistic dopings \cite{kp-review}. 
The situation can be more complicated for  n-doped samples, especially 
for WS$_2$ and WSe$_2$. For these two materials the states in the six $Q$ valleys are likely to 
be nearly degenerate with the states in the $\pm K$ valleys. Therefore the $Q$ valleys might be 
relatively easily populated for finite n-doping and, in contrast to the $\Gamma$ point, the 
effective mass is comparable to that  in the $\pm K$ \cite{kp-review} valleys. 
Therefore they may  contribute to the SdH oscillations. They would also affect the level 
broadening of the  $\pm K$ valley states because scattering from $K$ ($-K$) to 
three of the six  $Q$ valleys is not forbidden by spin selection rules \cite{kp-review}. 
Furthermore, we note that in the absence of a magnetic field 
the six $Q$ valleys are pairwise connected by time reversal symmetry. Therefore, 
taking into account only the lowest-in-energy spin-split band in the  $Q$ valleys, 
the  LLs belonging to the  $Q$ valleys will be three-fold degenerate: the magnetic field,
similarly to the case of the $K$ and $-K$ points, would lift the six-fold valley-degeneracy. 
The effective valley  g-factors, however, might be rather different from the ones in the 
$\pm K$ valleys. 
For n-doped monolayer MoX$_2$ materials 
the situation is probably less complicated because the $Q$ valleys are higher in 
energy and are not as easily populated as for the WX$_2$ monolayers. For MoS$_2$ 
monolayers, therefore, one can neglect the $Q$ valleys in first approximation.


\section{Summary}

In summary, we have studied the LL spectrum of monolayer TMDCs in a 
$\mathbf{k}\cdot\mathbf{p}$ theory 
framework. We have shown that in a wide magnetic field range the effects of 
the trigonal warping in the
band structure are not very important for the LL spectrum. Therefore the LL spectrum 
can be approximated by a harmonic oscillator spectrum and a linear-in-magntic 
field term which describes 
the VDB.  This approximation and the assumption that only intra-valley intra-band 
scattering is relevant  allowed 
us to extend previous theoretical work on SdH oscillations to the case of monolayer TMDCs. 
In the semiclassical limit, where analytical calculations are possible, 
it is found tha the VDB affects the amplitude of the SdH oscillations, 
whereas the spin-splitting of the bands leads to a phase difference in the oscillatory components. 
Since in  actual experimental situations  there might be only a few occupied LL below $E_F$, we 
have also performed numerical calculations for the conductance oscillations and  compared them 
to the analytical results. As it can be expected, if there are only a few LLs  populated 
the amplitude of the SdH oscillations obtained in the semiclassical limit does not agree 
very well with the 
results of numerical calculations.  This should be taken into account in the analysis of the 
experimental measurements. We used our theoretical results to analyze the measured SdH oscillations 
of Reference \cite{hone-SdH}. It is found that the quantum scattering time relevant for the 
SdH oscillations is significantly shorter than the transport scattering time that can 
be extracted from
the Hall mobility. Finally, we briefly discussed the effect of other valleys in the 
band structure on the SdH oscillations.


\section{Acknowledgement}

A.K. thanks Xu Cui for discussions and for sending some of their experimental data 
prior to publication. 
This work was supported by Deutsche Forschungsgemeinschaft (DFG) 
through SFB767 and the European Union through 
the Marie Curie ITN S3NANO. P. R. would like to acknowledge the support of the 
Hungarian Scientific Research Funds (OTKA) K108676.




\section*{References}


\begin{thebibliography}{99}


\bibitem{nature-phys-review}
Xu X, Yao W, Xiao D, and Heinz T F,
2014 \textit{Nature Physics} \textbf{10}, 343. 

\bibitem{chemsoc-review}
Liu G-B, Xiao D, Yao Y, Xu X, and Yao W, 
2015 \textit{Chem.\ Soc.\ Rev.}, Advance Article


\bibitem{kp-review} 
Korm\'anyos A, Burkard G, Gmitra M, Fabian J, Z\'olyomi V, Drummond N D, and Fal'ko V I, 
2015 \textit{2D Materials}  \textbf{2}, 022001.




\bibitem{heinz2010}
Mak K F, Lee C, Hone J,  Shan J,  and Heinz T F,
2010 \textit{Phys.\ Rev.\ Lett.}\ \textbf{105},   136805.



\bibitem{dxiao}
Xiao D,  Liu G B, Feng W,  Xu X, and Yao W, 
2012 \textit{Phys.\ Rev.\ Lett.}\  \textbf{108},  196802.


\bibitem{heinz2012}
Mak K F, He K,  Shan J,  and Heinz T F, 
2012 \textit{Nature Nanotechnology} \textbf{7}, 494.



\bibitem{cui2012}
Zeng H, Dai J, Yao W, Xiao D, and Cui X,
2012 \textit{Nature Nanotechnology} \textbf{7}, 490. 



\bibitem{cao}
Cao T, Wang G, Han W, Ye H, Zhu Ch, Shi J, Niu Q,
Tan P, Wang E, Liu B, and Feng J, 
2012 \textit{Nature Communications} \textbf{3}, 887. 


\bibitem{sallen}
Sallen G, Bouet L, Marie X,  Wang G, Zhu C R, Han W P, Lu Y,
Tan P H, Amand T, Liu B L, and Urbaszek B,
2012 \textit{Phys.\ Rev.\ B} \textbf{86},  081301.


\bibitem{korn}
 Korn T, Heydrich S, Hirmer M, Schmutzler J, and Sch\"uller C,
 2011 \textit{Appl.\ Phys.\ Lett.}\ \textbf{99}, 102109.  

 
 \bibitem{jones}
Jones A M, Yu H, Ghimire N, Wu S, Aivazian G, 
Ross J S, Zhao B, Yan J, Mandrus D, Xiao D, Yao W,  and Xu X, 
2013 \textit {Nature Nanotechnology} \textbf{8}, 634. 
 
 
\bibitem{gedik}
Sie E J, McIver J W, Lee Y-H, Frenzel A J, Fu L, Kong J, and Gedik N,
2014 \textit{Nature Materials} \textbf{14}, 290.
 
 
 
\bibitem{urbaszek2015a}
Wang G, Marie X, Gerber I, Amand T, Lagarde D, Bouet L, Vidal M, 
Balocchi A, and Urbaszek B,
2015 \textit{Phys. Rev. Lett.}  \textbf{114}, 097403.
 

\bibitem{kis-transistor1}
Radisavljevi\'c B,  Radenovi\'c A,  Brivio J,  Giacometti V, Kis, A, 
2011 \textit {Nature Nanotechnology} \textbf{6}, 147. 


\bibitem{jarillo-herrero2015}
Wang J , Yang Y, Chen Y-A, Watanabe K, Taniguchi T, Churchill H O H, and Jarillo-Herrero P, 
2015 \textit{Nano Letters} \textbf{15}, 1898. 


\bibitem{chhowalla}
Kappera R,  Voiry D, Yalcin S E, Branch B, Gupta G, Mohite A D, and Chhowalla M,
2014 \textit{Nature Materials} \textbf{13}, 1128.




\bibitem{zhixian}
Chuang H-J, Tan X,  Ghimire N J,  Perera M M, 
Chamlagain B,  Cheng M M-Ch, Yan J, Mandrus D, Tom\'anek D, and Zhou Z,
2014 \textit{Nano Letters} \textbf{14}, 3594.


\bibitem{hone-SdH}
Cui X, Lee G-H, Kim Y D, Arefe G, Huang  P Y, Lee Ch-H, Chenet D A, Zhang X, Wang  L, Ye F, 
Pizzocchero  F, Jessen  B S, Watanabe  K, Taniguchi  T, Muller  D A, Low T, Kim P, and Hone  J,
2015 \textit{Nature Nanotechnology}, \textbf{10} 534. 
(see also \textit{Preprint} arXiv:1412.5977).



\bibitem{duan-graphene-electr}
Liu Y, Wu H, Cheng H C, Yang S, Zhu E, He Q, Ding M, Li D, Guo J, Weiss N O, Huang Y, 
and Duan X, 
2015 \textit{Nano Letters} \textbf{15} 3030. 



\bibitem{ningwang-SdH}
Xu Sh, Han Y, Long G, Wu Z, Chen X, Han T, Ye W, Lu H, Wu Y, 
Lin J, Shen J, Cai Y, He Y, Lortz R, and Ning Wang,  
2015 \textit{Preprint} arXiv:1503.08427. 



\bibitem{radisavljevic2013}
Radisavljevic B and Kis A, 
2013  \textit{Nature Materials} \textbf{12} 815.


\bibitem{jarillo-herrero2013}
Baugher B W H, Churchill H O H, Yang Y, and Jarillo-Herrero P,
2013 \textit{Nano Letters} \textbf{13},  4212.


\bibitem{peide-ye}
Neal A T, Liu H, Gu J, and Ye P D, 
2013 \textit{ACS Nano} \textbf{8}, 7077. 


\bibitem{goki-eda}
Schmidt H, Yudhistira I, Chu L, Castro Neto A H, \"Ozyilmaz B, 
Adam S, and Eda G, 
2015 \textit{Preprint} arXiv:1503.00428





\bibitem{shen}
Lu H-Zh, Yao W, Xiao D and Shen Sh-Q, 
2013 \textit{Phys. Rev. Lett.} \textbf{110}, 016806. 



\bibitem{OchoaFalko}
Ochoa H, Finocchiaro F, Guinea F, and Fal'ko V I, 
 2014 \textit{Phys.\ Rev.\ B} \textbf{90}, 235429.




\bibitem{chuanwei-zhang}
Chu R-L, Li X, Wu S, Niu Q, Yao W, Xu X, and Zhang Ch,
2014 \textit{Phys. Rev. B} \textbf{90}, 045427. 



\bibitem{ho}
Ho Y-H, Chiu Ch-W, Su W-P, and Lin M-F, 
2014 \textit{Appl. Phys. Lett.} \textbf{105}, 222411. 

 
\bibitem{asgari2015}
Rostami H and Asgari R,
2015 \textit{Phys. Rev. B} \textbf{91} 075433.


\bibitem{niu2013a}
Li X, Zhang F, and Niu Q,
2013 \textit{Phys. Rev. Lett.} \textbf{110}, 066803.
 


\bibitem{niu2013b}
Cai T, Yang S A, Li X, Zhang F, Shi J, Yao W, and Niu Q,
2013 \textit{Phys. Rev. B} \textbf{88}, 115140. 


\bibitem{rose}
Rose F, Goerbig M O,  and  Pi\'echon F, 
2013 \textit{Phys. Rev. B} \textbf{88},  125438.




\bibitem{asgari2013}
Rostami H,  Moghaddam A G, and Asgari R,
2013 \textit{Phys.\ Rev.\ B} \textbf{88},  085440.




\bibitem{kormanyos-prx}
Korm\'anyos A, Z\'olyomi V, Drummond N D, and Burkard G,
2014 \textit{Phys.\ Rev.\ X}  \textbf{4}, 011034.



 
 


\bibitem{qiu}
Qiu D Y, da Jornada F H, and  Louie S G, 
2013 \textit{Phys.\ Rev.\ Lett.}\ \textbf{111}, 216805.

\bibitem{ashwin}
Ramasubramaniam A, 
2012 \textit{Phys.\ Rev.\ B} \textbf{86},  115409.



\bibitem{macneill}
MacNeill D, Heikes C, Mak K F,  Anderson Z, Korm\'anyos A, Z\'olyomi V, Park J, and  Ralph D C,
2015 \textit{Physical Review Letters} \textbf{114}, 037401.



\bibitem{urbaszek2015b}
Wang G, Bouet L, Glazov M M, Amand T, Ivchenko E I, Palleau E, Marie X, and Urbaszek B,
2015 \textit{2D Materials} \textbf{2} 034002.



\bibitem{holstein}
Adams E N, and Holstein T D, 
1959 \textit{J. Phys. Chem. Solids} \textbf{10}, 254. 
 

\bibitem{gzhou}
Zhou X, Liu Y, Zhou M, Tang D, and Zhou G,
2014 \textit{J. Phys.: Condens Matter} \textbf{26}, 485008.


\bibitem{leroy2015}
Yankowitz M,  McKenzie D,  and  LeRoy B J,
2015 \textit{Phys. Rev. Lett.} \textbf{115}, 136803.
 



\bibitem{ando}
Ando T, 
1974 \textit{J. Phys. Soc. Jpn.} \textbf{37}, 1233.



\bibitem{raichev-book}
Vasko F T and Raichev O E, 
\textit{Quantum Kinetic Theory and Applications}, Springer (2005). 




\bibitem{flensberg-book}
Bruus H and Flensberg K, \emph{Many-Body Quantum Theory in Condensed Matter Physics}, 
Oxford University Press (2006). 



\bibitem{raikh}
Raikh M R, and Shahbazyan T V,
1994 \textit{Phys. Rev. B} \textbf{49}, 5531. 


\bibitem{willander}
Averkiev N S, Golub L E, Tarasenko S A, and Willander M, 
2001 \textit{J. Phys.: Condens. Matter} \textbf{13}, 2517.


\bibitem{raichev}
Raichev O E, 
2008 \textit{Phys. Rev B} \textbf{78}, 125304. 




\bibitem{goerbig}
Goerbig M O, Montambaux G, and Pi\'echon F,
2014 \textit{EPL} \textbf{105} 57005. 




 
\bibitem{osgood2013}
Jin W,  Yeh P-CH,  Zaki N, Zhang D,  Sadowski J T,  Al-Mahboob A,
 van der Zande A M, Chenet D A,  Dadap J I,  Herman I P,
 Sutter P, Hone J, and  Osgood, R M Jr.,
2013 \textit{Phys.\ Rev.\ Lett.}\ \textbf{111}, 106801.

 
 



%
%
%





\end{thebibliography}
\end{document}